\documentclass{aa}  
\usepackage{natbib}
\bibpunct{(}{)}{;}{a}{}{,}
\usepackage{graphicx}
\usepackage{txfonts}
\def\simle{\mathrel{\hbox{\rlap{\hbox{\lower4pt\hbox{$\sim$}}}\hbox{$<$}}}}
\def\simgr{\mathrel{\hbox{\rlap{\hbox{\lower4pt\hbox{$\sim$}}}\hbox{$>$}}}}
\begin{document}

\title{$\gamma$ Cas stars: Normal Be stars with disks impacted by \\ the wind of a helium-star companion?}

\author{N.\ Langer\inst{1,2}
        \and
        D.\ Baade\inst{3}
        \and
        J.\ Bodensteiner\inst{4}
        \and
        J.\ Greiner\inst{5}
        \and
        Th.\,Rivinius\inst{6}
        \and 
        Ch.\,Martayan\inst{6}
        \and
        C.C.\,Borre\inst{7}}

   \institute{Argelander Institut f\"ur Astronomie der Universit\"at Bonn, Auf dem H\"ugel 71, 53121 Bonn, Germany\\
              \email{nlanger@astro.uni-bonn.de}
         \and
         Max-Planck-Institut f\"ur Radioastronomie, Auf dem H\"ugel 69, 53121 Bonn, Germany
         \and
             European Organisation for Astronomical Research in the Southern Hemisphere (ESO), Karl-Schwarzschild-Str.\ 2, 85748 Garching b.\ M\"unchen, Germany \\
             \email{dbaade@eso.org}
         \and
             Instituut voor Sterrenkunde, KU Leuven, Celestijnenlaan 200D, Bus 2401, 3001 Leuven, Belgium \\
             \email{julia.bodensteiner@kuleuven.be}
         \and
             Max-Planck-Institut f\"ur extraterrestrische Physik, Giessenbachstr.\ 1, 85748 Garching b.\ M\"unchen \\
             \email{jcg@mpe.mpg.de}
         \and
             European Organisation for Astronomical Research in the Southern Hemisphere (ESO), Casilla 19001, Santiago, Chile \\
             \email{triviniu@eso.org, cmartaya@eso.org}
         \and
             Stellar Astrophysics Centre, Aarhus University, Ny Munkegade 120, 8000 Aarhus C, Denmark \\
             \email{cborre@phys.au.dk}
}

\date{Received; accepted}
 
\abstract{
$\gamma$\,Cas stars are a $\sim$1\% minority among classical Be stars with hard ($\geq$5-10\,keV) but only moderately strong continuous thermal X-ray flux and mostly very early-B spectral type.  The X-ray flux has been suggested to originate from matter accelerated via magnetic disk-star interaction, by a rapidly rotating neutron star (NS) companion via the propeller effect, or by accretion onto a white dwarf (WD) companion.
In view of the growing number of identified $\gamma$\,Cas stars and the 
only imperfect matches between these suggestions and the observations, 
alternative models should be pursued. 
Two of the three best-observed $\gamma$\,Cas stars,
  \object{$\gamma$\,Cas} itself and \object{$\pi$\,Aqr}, have a
  low-mass companion with low optical flux; interferometry of
  \object{BZ\,Cru} is inconclusive.  Binary-evolution models are
  examined for their ability to produce such systems.
 The OB+He-star stage of
post-mass transfer binaries, which is otherwise observationally unaccounted, 
can potentially reproduce many observed properties of $\gamma$\,Cas\,stars.
The interaction of the fast wind of helium stars 
with the circumstellar disk and/or with the wind of Be stars 
may give rise to the production of hard X-rays.
While not modelling this process, it is shown that the energy budget
is favourable, and that the wind velocities may lead to
hard X-rays as observed in $\gamma$\,Cas
stars. Furthermore, the observed number of these objects appears 
to be consistent with the evolutionary models.
Within the Be+He-star binary model, the Be stars in $\gamma$-Cas stars are conventional 
classical Be stars.  They are encompassed by O-star+Wolf-Rayet systems 
towards higher mass, where no stable Be decretion disks exist, and 
by Be+sdO systems at lower mass where the sdO winds may be too weak to cause the 
$\gamma$\,Cas phenomenon.  
In decreasing order of the helium-star mass, the
descendants could be Be+black-hole, Be+NS or Be+WD binaries.
The interaction
between the helium-star wind and the disk may provide new
diagnostics of the outer disk.  }
\keywords{Stars: emission-line, Be -- stars: circumstellar matter --
  stars: evolution -- binaries: general -- X-rays: stars -- stars -
  individual: $\gamma$\,Cas, BZ\,Cru, $\pi$\,Aqr}
\maketitle
%

\section{Introduction}
\label{intro}
Many of the most rapidly rotating non-supergiant B, late O, and early
A stars exhibit H$\alpha$ line emission \citep{1997A&A...318..443Z,
  2001A&A...368..912Y}.  Typically, the emission lines form in a
Keplerian disk and the central stars rotate at very roughly 80\% of the
critical velocity \citep{2012A&A...538A.110M}.  Stars with these
properties are commonly called (classical) Be stars which were broadly reviewed by \citet{2013A&ARv..21...69R}.  The Be Star Spectra database \citep[BeSS, ][]{2011AJ....142..149N} lists nearly 250 Be stars with v\,$\leq$\,6.5\,mag.

\citet{1931ApJ....73...94S} suggested that the disks of Be stars are
the result of stellar rotational instability.  On the one hand, the
paucity of Be stars observed to rotate critically \citep{2013A&ARv..21...69R} appears to invalidate
this simple hypothesis as a general property of Be stars.  On the other hand, during the course of
stellar evolution, core contraction and envelope expansion would
combine to a net outward angular-momentum transport which, given a
sufficient initial supply \citep[as found by][]{2007A&A...462..683M},
could eventually lead to critical rotation \citep{2013A&A...553A..25G,
  2011A&A...530A.115B}.  As explained by \citet[][their Sect.\,5.2.5
and references therein]{2018MNRAS.476.3555R}, Be stars can avert the
possible permanent angular-momentum crisis by the viscous decretion of
matter and associated angular momentum.  Viscosity can enable the
formation of a Keplerian disk by redistributing the specific angular
momentum of ejected matter such that a $\sim$1\% fraction reaches
Keplerian velocities and the rest falls back to the star
\citep{1991MNRAS.250..432L}.  The variability of the mass content of
the disk may provide a means to estimate the amount of angular
momentum lost along with the matter \citep{2018MNRAS.479.2214G,
  2018MNRAS.476.3555R}.

An obvious alternative mechanism to spin up Be stars is mass transfer
in a binary.  In fact, in some classical Be stars hot subluminous
companions have been found \citep[][for other examples see
below]{2016ApJ...828...47P, 2018ApJ...853..156W} so that the high spin
rate of the B star may be the result of mass transfer from the
companion, which initially was the more massive star.  The
effectiveness of viscous decretion to build Keplerian disks is
unaffected by sufficiently distant companion stars.  Therefore, viscous
decretion is thought to be a universal property of Be
stars because Be stars with known short orbital periods are very rare (however, it is well possible that the frequency of such systems is reduced if the formation of stable decretion disks is hindered by the companion). For
viscous decretion being able to form Keplerian disks, it must be
supplied with matter by a stellar mass-loss process.  The ubiquity of
nonradial pulsations (NRPs) in Be stars \citep{2013A&ARv..21...69R,
  2017sbcs.conf..196B, 2018A&A...613A..70S} and the co-phasing of
apparent mass-loss events with maxima of the vectorial amplitude sum
of multiple pulsation modes \citep{2018pas8.conf...69B} suggest
strongly to search for the root of the mass loss in multi-mode
NRPs in single as well as binary Be stars.
Most probably, single- as well binary-star 
formation channels of Be stars are also realised by
nature, either alone or in combination.  This paper considers the
binary channel only, without implication for the single-star channel.

Since the X-ray luminosity of OB stars is proportional to their
bolometric luminosity \citep{2009A&ARv..17..309G}, X-rays due to
shocks in the winds are not an important property of isolated Be stars
\citep{1997ApJ...487..867C}.  A possible small X-ray excess in Be stars
w.r.t.\ normal B stars \citep{2000ASPC..214..156C} may be due to
additional shocks in the interface region between wind and disk.
However, some binary Be stars do reveal themselves through prominent, often
strongly modulated X-ray emission.  In the vast majority of these Be X-ray binaries
\citep[BeXRBs,][]{2011Ap&SS.332....1R}, a neutron star
accretes matter when it passes through or close to the Be star's
circumstellar disk, and part of the gravitational energy released in
the accretion process is emitted in the X-ray domain.  While the X-ray
flux of all BeXRB detected in early surveys is pulsed, systematic
searches in nearby galaxies are beginning to identify sources without
short periods \citep{2016A&A...586A..81H}; either they are genuinely
aperiodic, or the periods were not found because they are too long to
be easily determined.

A second subclass, which accounts for $\sim$1\% of all classical Be
stars, is also identified on the basis of X-ray properties
\citep{2016AdSpR..58..782S, 2018A&A...619A.148N}.  These stars emit
unusually hard ($\geq$5-10\,keV) but only moderately strong X-rays
which are variable on all timescales and distinctly thermal
\citep[][see also Sect.\,\ref{gCstars}]{2018A&A...619A.148N}.  While
the hardness is not too discrepant from observations of BeXRBs, the
X-ray luminosity of accreting BeXRBs is much higher.  The prototype of
this second subclass is \object{$\gamma$\,Cas}.  Accordingly, the
other members are often called $\gamma$\,Cas stars.  $\gamma$\,Cas is
also the first Be star that was discovered
\citep{1866AN.....68...63S}.  For this reason, $\gamma$\,Cas is
considered by some as the prototype of Be stars \citep[for instance
the General Catalog of Variable Stars,][calls all Be stars `GCAS' (or
'BE') stars]{2017ARep...61...80S}.  However, $\gamma$\,Cas has a number
of observed properties that only few Be stars share
\citep{2002ASPC..279..221H} although it is not clear to what extent this 
is due to the particularly rich database.  The most important 
difference is the mentioned X-ray flux.

Because the X-ray properties of $\gamma$\,Cas do not match any
conventional category of X-ray sources in early-type stars, Smith and
collaborators \citep[see][for references]{2016AdSpR..58..782S} have,
in a long series of papers, developed the unconventional notion that
the X-rays from $\gamma$\,Cas result from the interplay between two
magnetic fields, one at the stellar surface and the other in the disk.
Both are said to be not observationally detectable because of their
small spatial scales.  Nevertheless, there seems to be the associated
hope \citep{2015ApJ...806..177M} that this model, which in the
following will be called the magnetic model for short, may eventually
explain Be stars at large.  Because of the small fraction of Be stars
with $\gamma$\,Cas-like X-ray properties and the elusiveness of direct
observational evidence for the suggested magnetic fields, it is
important that no mistake is made with any extrapolating
generalisation.

This paper will develop a completely different concept to explain
$\gamma$\,Cas stars which has little implication for the majority of 
Be stars.  It incorporates without restriction the general picture that
has been sketched above of classical Be stars so that $\gamma$\,Cas
stars are ordinary Be stars with some additional properties.  The
proposed main difference is the response of the circumstellar Be disk
and/or the Be wind to the impact of a fast wind from a helium-star
companion.

For a better understanding, the key properties of the currently most
prominent $\gamma$\,Cas stars, namely $\gamma$\,Cas itself,
\object{$\pi$\,Aqr}, and \object{BZ\,Cru}, are recalled in
Sect.\,\ref{gCstars}.  Section\,\ref{models} describes the magnetic
model in more detail as well as the white-dwarf
\citep{2018PASJ...70..109T} and the magnetic-neutron-star propeller
model \citep{2017MNRAS.465L.119P} that were recently proposed as
alternatives.  Because $\gamma$\,Cas and $\pi$\,Aqr are binaries, the
role of binarity in Be stars is reviewed in the context of extent
observations (Sect.\,\ref{binobs}) and evolutionary models
(Sect.\,\ref{binmods}).  The conclusions are bundled in
Sect.\,\ref{summary}.

\section{Observed properties of $\gamma$\,Cas stars}
\label{gCstars}

\subsection{Overview}
At this moment, a Be star is admitted to the $\gamma$\,Cas family on
the basis of its X-ray flux if the latter is hard (L(2-10\,keV) /
L(0.5-2\,keV)\,$>$\,1.6), moderately strong
(log(L$_{\rm X}$/L$_{\rm bol}$)\,$\sim$\,--6), and thermal
\citep{2018A&A...619A.148N}.  These selection criteria have mostly
identified stars in the narrow spectral-type range of B0.5 to B1.5
(with luminosity classes V-III), although some exceptions are
beginning to be reported \citep{2018A&A...619A.148N}.  The 0.1-10\,keV
X-ray luminosity is intermediate between noninteracting Be stars and
BeXRBs.  Table\,\ref{gcasdat} reproduces the main properties of the
$\sim$15 currently known $\gamma$\,Cas stars as compiled by
\citet{2018A&A...619A.148N}.  A very useful account of the X-ray
properties of $\gamma$\,Cas stars and possibly related objects is
available from \citet{2018PASJ...70..109T}.

\begin{table*}
  \caption
  {Key observational data for the known $\gamma$\,Cas stars 
    taken from \citet[][their Table 5]{2018A&A...619A.148N}.  
    Soft and hard X-ray fluxes refer to the 0.5-2.0\,keV 
    and 2.0-10.0\,keV intervals, respectively.}
\label{gcasdat}
\centering
\begin{tabular}{r l l c c c c c c}
\hline\hline
No. &  Name         &  Spectral Type &  log($L_X/L_{bol}$) &   $L_X$         &  $L_{X,hard}$    & hardness ratio &    kT  & $v$\,sin\,$i$ \\
    &               &                &                    & 10$^{31}$ erg/s & 10$^{31}$ erg/s &                   &   keV  &  km/s          \\
\hline
1   & $\gamma$\,Cas & B0IV-Vpe       &     -5.39          &   85.0         &  65.1           &  3.25          & 14–25  &  295   \\
3   & V782 Cas      & B2.5III:[n]e+  &     -5.25          &   30.3         &  29.9           & 63.1           &  7     &        \\
17  & PZ Gem(high)  & O9pe           &     -6.14          &   9.66         &  7.87           &  4.32          & 16     &  265   \\
26  & HD90563       & B2Ve           &     -5.85          &   32.0         &                 &                &        &        \\
34  & BZ Cru        & B0.5IVpe       &     -5.69          &   27.6         &  20.3           &  2.81          & 13     &  338   \\
37  & HD119682      & B0Ve           &     -5.63          &   66.7         &  47.9           &  2.55          & 8-17   &  200   \\
39  & V767 Cen      & B2Ve           &     -5.37          &   26.2         &  17.4           &  1.97          &  6     &  100   \\
40  & CQ Cir        & B1Ve           &     -4.30          &   175          &  147            &  5.26          &  9     &  335   \\
47  & V759 Ara      & B2Vne          &     -5.29          &   41.9         &  31.9           &  3.21          & 10     &  277   \\
51  & V3892 Sgr     & Oe             &     -5.78          &   30.8         &  21.2           &  2.24          & 7-14   &  260   \\
53  & V771 Sgr      & B3/5ne         &     -4.64          &   24.5         &  21.1           &  6.29          &  8     &        \\
54  & HD316568      & B2IVpe         &     -6.26          &   4.04         &  2.43           &  1.60          & 4-6    &        \\
75  & V2156Cyg      & B1.5nnpe       &     -5.30          &   7.53         &  6.51           &  6.34          &  3     &        \\
79  & $\pi$ Aqr     & B1Ve           &     -5.59          &   7.44         &  5.80           &  3.56          & 12     &  243   \\
83  & V810 Cas      & B1npe          &     -5.14          &   48.4         &  41.1           &  5.58          &  64    &  422   \\
\hline
\end{tabular}
\end{table*}

The individual characteristics of the three best-observed
representatives are outlined in the following subsections.

\subsection{$\gamma$\,Cas}
\label{gCas}

After the first detection of X-rays from $\gamma$\,Cas
\citep{1976IAUC.2900R...1J}, there was not much of an alternative to a
classification as a BeXRB.  However, the lack of pulsing
\citep{1993A&A...275..227P} and regularly repeating X-ray outbursts
when a putative compact companion would in its (eccentric) orbit 
accrete matter from the Be disk \citep{2001A&A...377..161O} cast 
doubts on the origin of the X-rays, and the
nature of $\gamma$\,Cas (=\,HR\,264\,=\,HD\,5394\,=\,HIP\,4427) has
been controversial ever since.  

$\gamma$\,Cas was also one of the
first Be stars in which discrete absorption components (DACs) of UV
resonance lines were discovered \citep{1983ApJ...268..807H}.  DACs are
nearly universal in luminous OB stars \citep{1989ApJS...69..527H} and
usually attributed to corotating interaction regions in the wind that
originate from the high intrinsic instability of the wind, perhaps
triggered by photospheric inhomogeneities \citep{1996ApJ...462..469C}.
The azimuthal propagation of the interaction regions may lead to a
modulation of X-ray flux resulting from shocks in the wind
\citep{2001A&A...378L..21O}.  Because of their ubiquity in luminous
stars with radiatively driven winds, the DACs in $\gamma$\,Cas do not
reveal anything specific about the properties of this star, except
that its mass-loss process and wind are perfectly normal for an
early-type Be star.

Four periods have been reported for $\gamma$\,Cas and used in various
attempts to identify the nature of this star's X-ray activity.  The
orbital period of $\sim$203.5\,d was first identified by
\citet{2000A&A...364L..85H}.  Later refinements revised the
eccentricity to $\sim$0 and are based mainly on radial-velocity
measurements of the flanks of the H$\alpha$ emission-line profiles
\citep{2002PASP..114.1226M, 2012A&A...537A..59N}.  Although major
long-term corrections are required and the radial velocity of the disk
is not the same as that of one of the two stars, neither the value of
the period nor its nature are disputed.  The orbital period was also
found in the temporarily flat top of the H$\alpha$ emission-line
profile (Borre et al., in prep.), which is probably orbitally
modulated by the interaction of the companion with the (spiral) disk
structure \citep[{cf.\,}][]{2018MNRAS.473.3039P}.
\citet{2000A&A...364L..85H} propose a likely mass range of the primary\footnote{
With {primary}, we designate the brighter of the two stars in a binary.}
between 13 and 18 M$_{\odot}$.

The nature of the companion is not well constrained.  The mass is
about one solar unit, and \citet{2012A&A...537A..59N} suggested that
it might be a helium star.  \citet{2002PASP..114.1226M} find
inhomogeneities in the disk and consider as one possible explanation
that H$\alpha$-emitting material is associated with the secondary.  In
search for a spectral signature of the secondary,
\citet{2017ApJ...843...60W} cross-correlated the UV spectrum with
model sdO spectra.  However, this effort failed because the very hot
primary dominated the cross-correlation function which, moreover, is
very broad due to the rapid rotation of the B-star primary.  Probably because
of the unfavourable magnitude difference at the wavelengths used,
long-baseline H$\alpha$ \citep{2006AJ....131.2710T} and K-band
\citep{2007ApJ...654..527G} interferometry has not detected the
companion either.  However, the circumstellar disk was resolved and
and the derived inclination angles of 55$^\circ$ and 51$^\circ$, respectively, are in very good agreement.

For 15 years, a 1.216-d period was seen in single-site groundbased
photometry \citep{2012ApJ...760...10H} but eventually dropped below
the detection threshold of very few mmag.  Both the frequency and the
decay in amplitude of this second periodic variability were also found
in SMEI space photometry (Borre et al., in prep.).  Later space
photometry with BRITE-Constellation confirmed the absence of the
1.216-d period at the 2-3-mmag level \citep[][Borre et al., in
prep.]{2017sbcs.conf..196B}.  Instead, BRITE detected a very nearly,
but probably not perfectly, three times shorter third period at
0.403\,d (frequency: 2.48\,c/d) with a peak-to-peak amplitude slowly
varying between $\sim$2 and $\sim$9\,mmag \citep[][Borre et al., in
prep.]{2017sbcs.conf..196B}.  A fourth frequency was identified at 1.25\,c/d 
in the SMEI observations (Borre et al., in prep.).  An attempt was made to use Doppler
shifts of the 2.5-c/d frequency to locate the site of the variability
in the system.  However, the time baseline of the BRITE data was too
short, and the systematic noise of the SMEI observations was too large
(Borre et al., in prep.).

The long-term constancy of the three short periods implies either
rotation or pulsation as their origin.  Rotationally induced
variability with period $P$ would require some physical property to
vary along the star's circumference with an azimuthal scale of
$(P/P_{\rm rot}) \times 2\pi$.  For instance, temperature, abundances or
magnetic structures.  There is no such report for $\gamma$\,Cas (apart
from the optical broad-band flux).  Radial pulsations are not known in
Be stars but both short periods are well within the range of 
NRPs found in other Be stars \citep{2016A&A...593A.106R,
  2017sbcs.conf..196B, 2018A&A...613A..70S}.  Since the 1.216\,d
variability faded while the 0.403\,d variability rose, it is plausible
to believe that both are of the same nature, which can, then, only be
NRPs.  In fact, space photometry \citep{2016A&A...593A.106R,
  2017sbcs.conf..196B, 2018A&A...613A..70S} has detected multiple
low-order NRP modes in many Be stars over the full range of B-type
stars.

Ongoing and forthcoming large-scale photometric surveys from space
will show how typical (multi-mode) NRP is for Be stars.  If the
pulsation properties of Be stars are different from those of Bn stars
(very rapidly rotating B-type stars identified through their
equator-on orientation but not known to have exhibited emission lines,
i.e., not possessing a circumstellar disk), this would be a strong
indicator that NRPs are a defining property of Be stars, probably
through their involvement in mass-loss events feeding the disk.

Additional periods may be hidden in complex spectroscopic line-profile
variability.  In agreement with quite similar observations in other
early-type stars, \citet{1988PASP..100..233Y} and
\citet{1994PASJ...46....9H} attributed such variability in optical
absorption lines also of $\gamma$ Cas to high-order NRP.
Intermediate- to higher-order NRP modes were also deduced from long
series of spectra of other Be stars
\citep[e.g.,][]{1993ApJ...417..320R, 1997ApJ...481..406K}, including
$\pi$ Aqr \citep{2005ASPC..337..294P}.  \citet{2016AdSpR..58..782S}
rejected the NRP hypothesis for $\gamma$\,Cas because they found the
variations of UV lines to be erratic and each migrating subfeature in
the line profiles to maintain its identity for no more than very few
hours.

However, the 30 hours, i.e.\, only about one
rotational period, of HST spectroscopy considered by
\citet{2016AdSpR..58..782S} are without doubt insufficient for the
proper tracking of features with similar but different propagation rates and for
the determination of their periods.  Accordingly, the suggestion by
\citet{1998ApJ...507..945S} that the subfeatures are only
rotationally advected is lacking a solid observational foundation.
By contrast, \citet{2005ApJ...623L.145W} observed
\object{$\zeta$\,Oph} (O9.5\,Ve) for 24 days with the {\it MOST}
space photometer and during 17 of these 24 days with three
spectrographs at different geographical longitudes.  They detected
at least a dozen photometric and eight spectroscopic periods.  Six
periods were in common to both datasets and interpreted as
intermediate-order NRPs.  An obvious rotation period was not
identified, and the multi-periodicity of the migrating subfeatures
rules out the rotational hypothesis for them.

In addition to the three genuine periods in $\gamma$\,Cas, there are
also cyclic optical broad-band flux variations on seasonally changing
timescales around 70\,d with a total range of 50-91\,d.  The
peak-to-peak amplitude of $\sim0.02$\,mag is not too far from the
sensitivity threshold to so slow variations of single-site groundbased
photometry.  \citet{2002ApJ...575..435R} combined the earlier cycles
into a single sinusoid with adaptive period and compared this
variable-stretch pseudo-sine curve of optical light to the X-ray flux.
They derived a correlation in the variability of the two domains using
only two photometric seasons and just six epochs of X-ray data.  Since
the pseudo-sine curve interpolates the light curve, the effective
comparison is between seasonally fragmentary optical-flux and very
patchy X-ray observations.  There is no assurance that such a data
treatment can lead to a stress-resistant conclusion.

From just one day of simultaneous X-ray and UV observations, Smith and
collaborators \citep[for references see][]{2016AdSpR..58..782S}
inferred correlations between X-ray flux on the one hand and UV flux,
UV spectral lines, etc. on the other.  However, it is not clear
that coincidences of two features each in two short datasets can carry
high weight in an object that in all observed wavelength regions is
variable on all timescales.  More significant is the correlation over
15 years between X-ray and optical flux reported by
\citet{2015ApJ...806..177M} although it is not clear which effect the
choice of the time windows has.  From their comparison, these authors
conclude that the X-rays lag the optical flux by no more than a month.  
Because the radial drift velocity of matter in Be disks is only of 
the order of a few km/s \citep{1999A&A...348..831R}, 
the time delay of X-ray emission due to accretion by a companion at an 
au-scale distance would be much longer.  By contrast, 
a lag by only a month is more plausible if it takes a month for the disk 
to build up and the interaction between the two postulated magnetic fields to commence.

However,  in the cross-correlation function, there is a broad and not
well separated peak near three years.  In view of this network of
claimed correlations, it surprises that the purported rotation period
has not been seen modulating any observable (other than the optical
flux).

\citet{2010A&A...512A..22L} emphasised the need for multi-component
fits of the continuum X-ray flux distribution.  From
high-spectral-resolution XMM-Newton observations with a complex
emission-line spectrum, they derived optically thin thermal emissions
at four discrete temperatures, namely 12-14\,keV, perhaps at
2.4\,keV, and with confidence at 0.6 and 0.11\,keV.  From observations
between 0.6 and 100\,keV, \citet{2015ApJ...799...84S} firmly rule out
any power-law component and thereby confirm the thermal nature of the
X-ray flux.  \citet{2012A&A...540A..53S} report that after an apparent
mass-loss event (ejection of matter into the disk), an absorbing layer
developed temporarily, indicating the presence of additional matter
along the line of sight.  Temperature contrasts are also evidenced by
spectral lines \citep{2010A&A...512A..22L, 2012A&A...540A..53S}.
 
Using independent observations, \citet{2018PASJ...70..109T} basically
agree with the stated decomposition of the X-ray continuum.  They also
confirm that changes in the hardness ratio are only weakly coupled to
flux variations, which mainly occur in the hottest plasma above 4\,keV
while the softer X-rays are more stable and are most of the time only
negligibly absorbed.  A new finding though are dips in softness,
especially of the ratio [0.5-2\,keV]/[4-9\,keV], which last a few ks.
Because these dips are unrelated to flux increases in the hard band,
\citet{2018PASJ...70..109T} conclude that these fadings are caused by
absorption in temporarily intervening matter.  This is consistent with
the similar picture derived by \citet{2012A&A...540A..53S} from X-ray
observations during an outburst of the B star.  Adopting the outburst
interpretation, it seems plausible that the X-ray-emitting volume was
(partly) located behind the ejecta.  The implied proximity to the B
star would argue, as may be deduced from the time delays between optical and X-ray fluxes, against the X-rays forming near a companion star at
an au-scale distance.

In the latest of his papers on $\gamma$\,Cas,
\citet{2019PASP..131d4201S} discusses various observations once again,
offering basically the same interpretations.  It seems useful to point
out that all the old observations were obtained with instruments not
employing solid-state detectors.  In those detectors, photons do not
merely excite electrons (internal photoelectric effect) but lead to
the physical emission of electrons (external photoelectric effect),
which are subsequently amplified and measured.  As the result,
measurements can in some cases deviate more from unbiased photon
statistics than is typical of solid-state detectors.  Moreover,
physically emitted electrons are more susceptible to subtle external
perturbations.

\subsection{$\pi$\,Aqr}
\label{piAqr}

On the basis of its X-ray properties, \citet{2017A&A...602L...5N}
recently classified \object{$\pi$\,Aqr}
(=\,HR\,8539\,=\,HD\,212571\,=\,HIP\,110672) as another $\gamma$\,Cas
star.  The similarity concerns not only the X-ray flux and hardness
but also the variability.  During the 50\,ks observations
with XMM-Newton, several
brightenings with a base width of 1-2\,ks occurred with pronounced
peaks reaching roughly thrice the previous or subsequent level. As in
$\gamma$\,Cas, no BeXRB-like outbursts have been observed.

After $\gamma$\,Cas itself and BZ\,Cru (Sect.\,\ref{bzCru}),
$\pi$\,Aqr became the third $\gamma$\,Cas star in the Bright Star
Catalog (and is moreover equatorial) so that also for $\pi$\,Aqr a
good record of its general properties and variability in other
wavelength regions is available.  \citet{2010ApJ...709.1306W}
documented the long-term stability of the disk
orientation in space by spectropolarimetry.  The decreasing H$\alpha$ emission strength
traced the dissipation of the disk over nearly a decade.  Variations
in H$\alpha$ equivalent width and continuum polarisation also caught a
number of outbursts \citep{2010ApJ...709.1306W} which are quite typical 
especially of early-type Be stars \citep{2018AJ....155...53L, 
2018MNRAS.479.2909B}.  The occurrence of DACs in UV wind lines
\citep{2006A&A...459..215S} is also common among Be stars
\citep{1989ApJ...339..403G}.

\citet{2002ApJ...573..812B} found that $\pi$\,Aqr is an 84.1-d binary.
The mass ratio is about 6:1, which should be more favourable for the
detection of the companion than the $\sim$15:1 ratio in $\gamma$\,Cas.
Depending on the inclination angle, the mass of the secondary may be
between 2.2 and 4.5\,M$_\odot$.  The orbital motion of the secondary
was derived from a 'travelling emission component', which the authors
attributed to a gaseous envelope surrounding the secondary.  From
H$\alpha$ profiles covering $\sim$40 orbits,
\citet{2013A&A...560A..30Z} extracted the same period for the
violet-to-red ratio $V/R$ of the two emission peaks.  Accordingly, the
disk structure is phase-locked to the position of the companion.  The
power spectrum plotted by \citet{2013A&A...560A..30Z} does not include
the first harmonic.  If this omission is justified, it would mean that
any two-armed spiral structure \citep{2018MNRAS.473.3039P} is not
axisymmetric, perhaps because one arm strongly dominates (or the two
arms are not 180 degrees apart in disk azimuth).  In fact, the study
identifies an extended region of enhanced H$\alpha$ line emission
between the two stars.  The strength of this emission follows the
long-term variability of the overall emission strength.  As for
$\gamma$\,Cas, the cross-correlation technique of
\citet{2017ApJ...843...60W} did not detect the companion to this hot
and broad-lined star.

\citet{2017A&A...602L...5N} put forward the argument that the
secondary in the $\pi$\,Aqr system is not a compact object itself and
that no such third body is likely to be in a closer orbit than the
secondary.  Therefore, they conclude that the X-ray properties of
$\pi$\,Aqr and, by implication, $\gamma$\,Cas stars in general are not
caused by a compact companion.  However, if the interaction of the
companion with the disk leads to additional H$\alpha$ emission
\citep[see also ][]{2002ApJ...573..812B}, more power seems required
than is available from an intermediate-mass main-sequence star.

As in $\gamma$\,Cas and several other Be stars, NRPs 
of intermediate degree ($m$=5) have been deduced from the photospheric
line-profile variability of $\pi$ Aqr \citep{2005ASPC..337..294P}.
\citet{2003A&A...411..229R} observed low-order line-profile
variability not matching the quadrupole NRP patterns typically seen in
Be stars.

\subsection{BZ\,Cru}
\label{bzCru}

The X-ray similarity to $\gamma$\,Cas of \object{BZ\,Cru}
(=\,HR\,4830\,=\,HD\,110432\,=\,HIP\,62027) was established by
\citet{2012ApJ...755...64S}.  In six visits by the {\it Rossi X-ray
  Timing Explorer (RXTE)}, each collecting 8-9 hours of observations
with the Proportional Counter Array \citep{1996SPIE.2808...59J}, well
over 1000 flares were seen by the authors.  With 5-s binning, they
could be as short 2.5 bins and lasted up to more than a minute with an
average rate of about one flare in twenty 5-s bins, i.e., not far from 
the confusion limit.  As in $\gamma$\,Cas, most of the time, the
hardness ratio did not change during the flaring.  On two occasions,
the X-ray emission subsided for a few hours.  The six datasets span
only 155 days; yet, the authors derived a 'period' of 226 days.
\citet{2018PASJ...70..109T} applied their models also to BZ\,Cru.  As
in the case of $\gamma$\,Cas, they achieved satisfactory fits of the
X-ray flux distribution but could not distinguish between a
nonmagnetic and a magnetic white dwarf.

As most other Be and supergiant OB stars with stellar winds, BZ\,Cru
exhibits variable DACs \citep{2012ApJ...755...64S}.  The same authors
also found intermittent migrating subfeatures in stellar line profiles
that, in other OB and Be stars, were attributed to nonradial
pulsation, but did not report periods.  They speculated about
``magnetically confined clouds'' but admitted that this is ``not
proven''.  As mentioned above (Sect.\,\ref{gCas}), such speculations were disproven in the
case of \object{$\zeta$\,Oph} \citep[O9.5\,Ve, ][]{2005ApJ...623L.145W}.

\citet{2006ApJ...640..491S} also noted that, if BZ\,Cru is a member of
NGC\,4609, it would be a blue straggler.  From a dedicated
interferometric search, \citet{2013A&A...550A..65S} only derived upper
detection limits for a companion star.  The disk had a strongly
asymmetric structure the nature of which could not be firmly
established.  \citet{2018ApJ...853..156W} did not detect the signature
of an sdO companion in {\it International Ultraviolet Explorer} ({\it
  IUE}) UV spectra.

\subsection{Synopsis}
\label{synopsis}

Any attempt to extract commonalities from a sample of just three,
albeit well-studied, representatives must appear presumptuous.
However, relying on \citet{2018A&A...619A.148N} for the X-ray
properties of $\gamma$\,Cas stars, the following working description is perhaps broadly
agreeable:
\begin{list}{--}{\itemsep=0mm\topsep=0mm}
\item
Typical spectral subtypes fall into the range B0.5 to B1.5.  
\item 
The X-ray flux is hard (L(2-10\,keV) / L(0.5-2\,keV)\,$>$\,1.6), 
moderately strong (log(L$_X$/L$_{bol}$)\,$\sim$\,--5.5), and thermal.
\item
The X-ray flux is variable on timescales from seconds to years.  
\item
Variations in the X-ray hardness ratio are small and mainly due to the 
hard component. 
\item
There are occasional reductions in the soft X-ray flux, consistent with 
intervening absorbers.
\item
At least on long timescales, X-ray and optical flux variations 
track each other. 
\item 
There is a lower-mass and optically faint companion (may
not be the case for BZ\,Cru).
\item 
The companion interacts with the Be disk.  (The not finally
explained strong asymmetry of the disk of BZ\,Cru may be caused by a
not otherwise detected companion.)
\item
Intervening absorbers ejected by the B star may localise the 
X-ray-forming region near the B star, not around the companion.  
\end{list}

\section{Current models for $\gamma$\,Cas stars}
\label{models}

\subsection{The magnetic model}
\label{magnet}

In addition to most of the observed properties listed in
Sect.\,\ref{synopsis}, the magnetic model rests on the assumptions
that (i) the 1.216-d period of $\gamma$\,Cas is the rotation period of
the B star, (ii) there is a correlation without major relative shift
in time between X-ray and optical flux, (iii) there is a correlation,
without offsets in time, between X-ray flux and spectral UV
features, (iv) the migrating subfeatures in spectroscopic line
profiles are not caused by nonradial pulsation, and (v) companion
stars are irrelevant for the understanding of $\gamma$\,Cas stars.  As
seen in Sect.\,\ref{gCstars}, all of these assumptions meet with
various degrees of doubt and cannot be proved or disproved using
currently available observations of $\gamma$\,Cas stars.

The magnetic model casts these assumptions into the notion of magnetic
fields as the common umbrella.  Two kinds of magnetic field are
envisioned.  One resides in the star and is said to arise from
sub-surface convection zones \citep{2011A&A...534A.140C}.  The other
one is pictured to result from the amplification by magneto-rotational
instability \citep[MRI; e.g.,][]{2000ApJ...543..486S} of seed fields
in the disk.  Circumstellar and stellar magnetic field lines are 
assumed to temporarily connect via fingers extending from
the disk towards the star.  Because of the different rotation rates of 
star and disk, the field lines are thought to be stretched, eventually 
disrupted and finally reconnected.  The snapping back of the field lines 
is suggested to accelerate charged particles to high
energies dissipated as X-rays when they hit the star.  Disk
instabilities and mass injections from the star are seen as the origin
of the assumed correlation, without much delay, between optical and
X-ray flux on the variable 70-d timescale.  Migrating subfeatures in
absorption lines are attributed to superphotospheric cloudlets forced
into corotation by the putative magnetic field
\citep{1998ApJ...507..945S}.  As discussed in Sect.\,\ref{gCas}, the
empirical basis for this latter belief is deficient.

According to \citet{2017MNRAS.469.1502S}, neither the stellar nor the
disk magnetic field postulated by them is directly observable because
their structures are thought to be too tangled and small-scale.
Therefore, the magnetic model is not a priori in direct conflict with
a survey of 85 Be stars \cite[incl.\ $\gamma$\,Cas, $\pi$\,Aqr, and BZ\,Cru] 
[, Neiner et al., in prep.]{2016ASPC..506..207W} which did not find one
star with a large-scale magnetic field whereas for non-Be early-type
stars in the same survey the typical fraction of magnetic stars is
about 5-10\% \citep{2016ASPC..506..207W}.  (A possible explanation of
this negative result is that a magnetic field would destroy a
Keplerian disk \citep{2018MNRAS.478.3049U} so that magnetic Be stars
cannot exist.  
By contrast, rapid rotation and a magnetic field are not strongly 
mutually exclusive for B0.5 to B1.5 stars near the zero age main sequence.)  
In spite of the lack of detected large-scale magnetic fields
in Be stars, the stellar magnetic field in
$\gamma$\,Cas stars is believed to be also responsible for the claimed
rotational modulation with extremely constant period of the optical
broad-band flux \citep{2016ASPC..506..215S}, which requires a
large-scale structure that does not migrate in the co-rotating frame.
The magnetic model does not address this obvious tension other than by
hypothesising that the non-detection of a magnetic field is related to
the disappearance of the photometric 1.215-d variability and due to
the decay of the stellar magnetic field \citep{2019PASP..131d4201S}.
If so, $\gamma$\,Cas would, during the presence of the 1.215-d period,
have possessed a large-scale magnetic structure not seen in any other
Be star.

The two magnetic constructs are imported from other contexts.
Convective sub-surface dynamos might produce variable inhomogeneous
surface brightness distributions which otherwise, in purely radiative
atmospheres, lack a simple explanation.  MRI is broadly invoked to
produce the level of viscosity needed to bring the timescales of
accretion processes into agreement with observational constraints
\citep{2019arXiv190101580M}.  (It is useful to note that also the
viscous-decretion-disk model for Be stars [cf.\ Sect.\,\ref{intro}] merely
assumes viscosity but does not explain it.)  However, to date, neither
magnetic-field-producing mechanism seems to have found direct
observational confirmation even in the domains they were designed for.
This motivates searches for alternate explanations of the
$\gamma$\,Cas stars.

The magnetic model was conceived before the (optically) faint and
low-mass companion star of $\gamma$\,Cas was discovered.  The model
has evolved over the years.  However, also in its current version it
does not foresee any role for the companion to contribute to the
observed phenomena.  This attitude is seemingly reinforced by the fact
that BeXRBs, where compact companions are the main X-ray actors, and
$\gamma$\,Cas stars are clearly distinguished populations.  If faint 
low-mass companions in low-eccentricity orbits are characteristic of 
$\gamma$\,Cas stars, a large difference in the X-ray properties 
of BeXRBs and $\gamma$\,Cas stars may be expected because in near-circular 
orbits the Be disk is strongly truncated.  As the result, the disk 
remains well within the B star's Roche lobe so that major 
X-ray outbursts with the orbital period are unlikely 
\citep{2001A&A...377..161O}.  Therefore, accreting binary models 
of $\gamma$\,Cas stars need to be powered by the B star's mass loss.

\subsection{Accretion onto a white dwarf}
\label{WD}

One way of avoiding the overproduction in $\gamma$\,Cas stars of
X-rays at the level of BeXRBs is to assume a white dwarf (WD) as the
accreting body because it has a shallower gravitational potential than
that of a neutron star or a black hole.  This was first proposed by
\citet{1995A&A...296..685H}.  In fact, accreting white dwarfs in novae
and symbiotic stars are X-ray sources of roughly comparable
properties.  Contrary to the purely parametric formalisms of most
earlier studies, \citet{2018PASJ...70..109T} employed models
specifically designed for white dwarfs accreting matter from a cool
companion as in novae or symbiotic stars and included reflection by
the white dwarf of X-rays as well as absorption.  They achieved
reasonable fits of the X-ray flux distributions of both $\gamma$\,Cas
and BZ\,Cru.  However, the models could not conclusively discriminate
between magnetic (as in polars or intermediate polars) and
non-magnetic (as in [dwarf] novae) WD companions.

\citet{2016ApJ...832..140H} offered the interesting idea that the
cooler X-ray emitting plasma ``probably originates from the Be primary
stellar wind, while the hot component may originate from the head-on
collision of either the Be or WD wind with the Be disk''.  In a
different context, it has, in fact, been shown that Be disks are
probably subject to ablation by the B star's radiation \citep[][and
references therein]{2018MNRAS.474..847K}.  However, if an interaction
between the wind from a Be star with the disk were at the origin of
the hard X-rays from $\gamma$\,Cas, more than just $\sim$1\% of the Be
stars should be $\gamma$\,Cas stars.  A variant of the suggestion of a
collision with a wind from a companion will be developed in
Sect.\,\ref{binmods}.

The mass estimate for the companion to $\gamma$\,Cas of one solar mass
is at the high end of WDs.  However, if the range of 2.2 to
4.5\,M$_{\odot}$ for the secondary star in $\pi$\,Aqr
\citep{2002ApJ...573..812B} is correct, the WD model would not be
applicable.  Depending on how much mass is transferred back to a WD
companion during the later evolution of the B-type primary and when
this happens, such systems might even be progenitors of a
thermonuclear Type Ia supernova explosion of the WD and a
core-collapse Type II supernova of the B star.

\subsection{The propeller model}
\label{propeller}

Recently, \citet{2017MNRAS.465L.119P} advanced the so-called propeller
model, which employs a neutron star but reduces the X-ray flux from
direct accretion as in BeXRBs by letting the magnetic field and rapid
rotation of the neutron star suitably moderate the accretion rate.
Moreover, because the X-ray emission is from a hot halo, it is not
rotationally pulsed (as observed).  This construct would seem to
eliminate the discrepancy in the X-ray domain between $\gamma$\,Cas
stars and BeXRBs.  However, \citet{2017MNRAS.469.1502S} have
nevertheless vehemently rejected the propeller model.  In particular,
they argue that the X-rays form close to the B star, not near the
companion at au-scale distances, because of intermittent X-ray
attenuations by cold plasma, ejected by the B star, between the
X-ray-emitting region and the observer.  Furthermore, they insist that
the density of the X-ray emitting plasma is of order
10$^{15}$\,cm$^{-3}$, i.e., at a photospheric level, while the
propeller model yields values near the inner radius of the
magnetosphere that are 1-2 orders of magnitude lower.

In addition, the assumption of neutron-star companions to
$\gamma$\,Cas and $\pi$\,Aqr is not straightforward.  Because BeXRBs
can exist for a few 10$^6$ years after the supernova explosion that
formed the neutron star whereas the remnant nebulae merge with the
interstellar medium within a few 10$^3$ years, the absence of such
nebulae around these stars is not an obstacle to the neutron-star
hypothesis.  Better indicators are, however, their orbital
eccentricity and the space velocity both of which may be
significantly modified by a supernova explosion.  This is briefly
discussed in the following two subsections.

\subsubsection{Impact of a supernova explosion on orbital eccentricity}

If a star exploding in a binary experiences a significant kick,
the eccentricity of the orbit grows, and the plane of the orbit may
get tilted with respect to the equatorial plane of the previous mass
gainer, which in the case of Be stars is also the plane of the disk.
The details depend very much on the direction of the kick
\citep{2019A&A...624A..66R}.  These expectations find their
confirmation in many observed BeXRBs \citep{2011Ap&SS.332....1R}.
They do not appear to be satisfied in $\gamma$\,Cas
\citep{2007ApJ...654..527G} and $\pi$\,Aqr \citep{2002ApJ...573..812B,
  2013A&A...560A..30Z} the orbits of which seem nearly circular.
\citet{2017MNRAS.465L.119P} invoke an electron-capture supernova
explosion for the progenitor of the assumed neutron star.  Such
explosions are thought to impart a low kick on the remnant.  However,
about 10\% of the rest mass of the exploding star is lost as
neutrinos.  Even if this mass loss is symmetric w.r.t.\ the center of
gravity of the exploding star, it is asymmetric about the binary's
center of gravity and so imposes some orbital eccentricity on binaries
that remain bound.

\subsubsection{Impact of a supernova explosion on space velocity}
\label{PM}

\begin{figure}
  \centering
  \includegraphics[width=7.8cm]{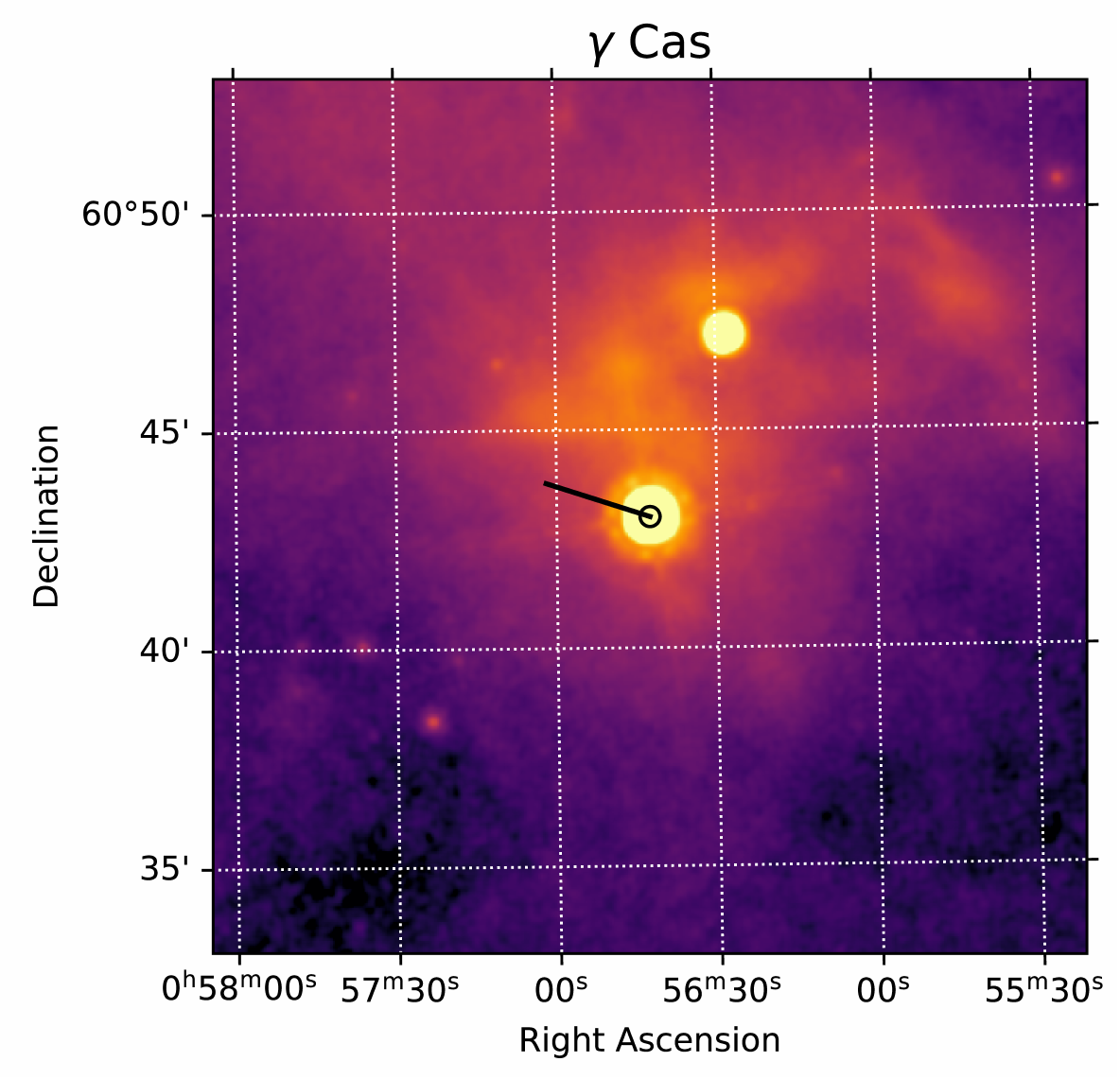}
  \caption{WISE 24$\mu$m image of $\gamma$\,Cas.  The black line at the
    center illustrates the 10,000-year proper motion (corrected for
    Galactic rotation) as measured by Hipparcos ($\gamma$\,Cas is too
    bright for Gaia DR2).}
  \label{gCasHip}
\end{figure}

\begin{figure}
  \centering
  \includegraphics[width=7.8cm]{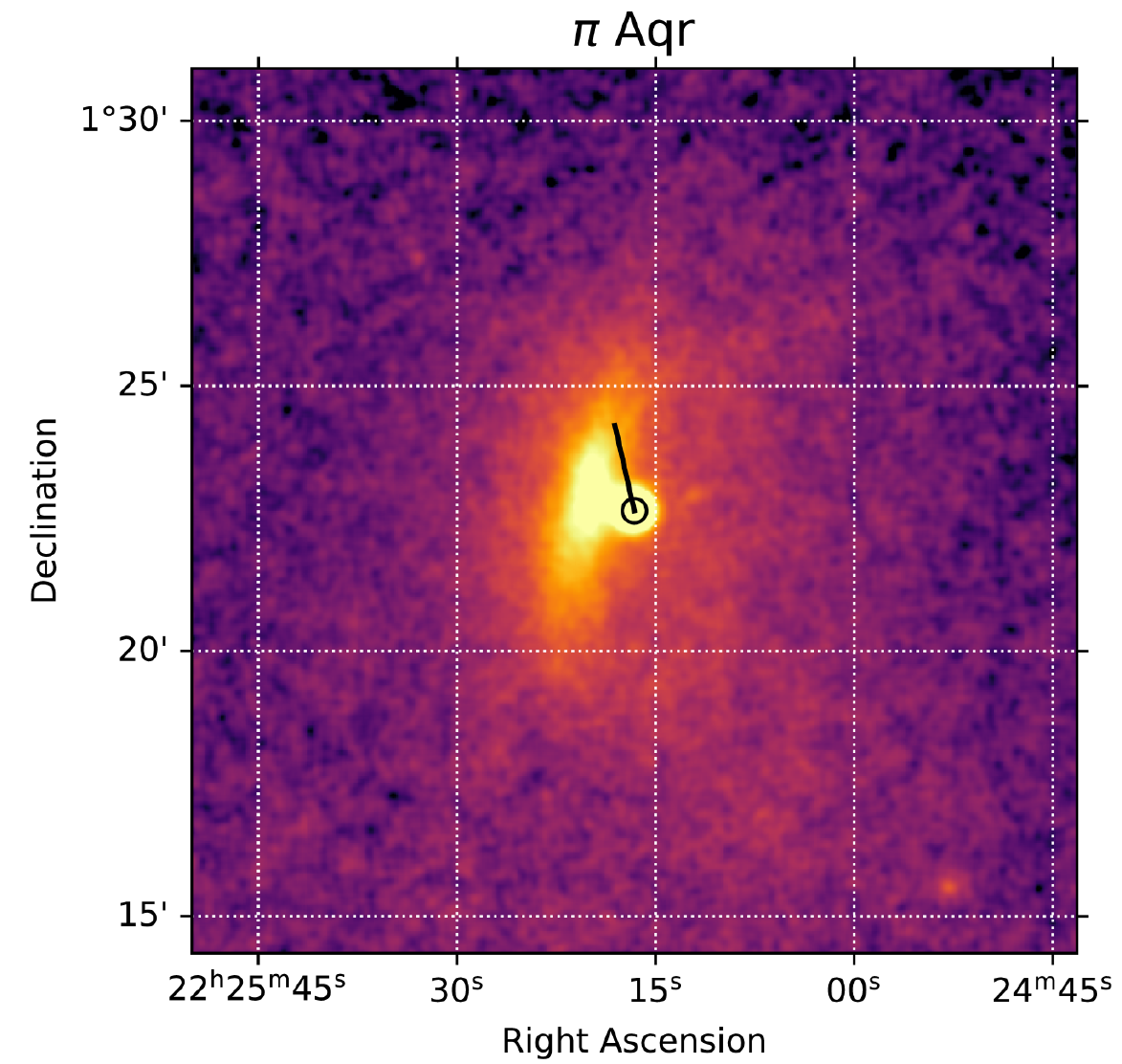}
  \caption{Ditto as Fig.\,\ref{gCasHip} except for $\pi$\,Aqr and a
    combination of the Hipparcos and Gaia proper-motion
    measurements.}
  \label{piAqrGaia}
\end{figure}

If a supernova explosion increases the velocity of a binary relative
to the ambient interstellar medium (ISM), a bow shock may develop when
a stellar wind impacts the ISM.  A prototypical case is the O9.5\,Ve
runaway star \object{$\zeta$\,Oph} \citep[][and references
therein]{2012A&A...543A..56D}.  However, as
\citet{2019A&A...624A..66R} explain, most surviving systems are not
expected to be accelerated by more than $\sim$30\,km/s.

\citet{2018A&A...618A.110B} have inspected and classified WISE
\citep{2010AJ....140.1868W} 24\,$\mu$m images of all OBA stars in the
Bright Star Catalog \citep{1991bsc..book.....H}, including
$\gamma$\,Cas and $\pi$\,Aqr.  WISE images of the regions around
$\gamma$\,Cas and $\pi$\,Aqr are reproduced in Figs.\,\ref{gCasHip}
and \ref{piAqrGaia}. The superimposed proper-motion vectors illustrate
the classifications by \citet{2018A&A...618A.110B} for $\gamma$\,Cas
and $\pi$\,Aqr, respectively.  A bow shock can be seen to be
associated with $\pi$\,Aqr \citep[see also][]{2016A&A...587A..30M}.
However, the apex of the nebula is not aligned with the proper-motion
vector.  Accordingly, the relative velocity of $\pi$\,Aqr and the
ambient interstellar medium is not dominated by the stellar motion.
$\gamma$\,Cas is also surrounded by a nebula, which may be related to
the star.  But the morphological evidence is weak so that the entry in
\citet{2018A&A...618A.110B} is ``not classified''.  Nevertheless, the
peculiar space velocities are close to ($\pi$\,Aqr: 21\,km/s) or even
well within \citep[$\gamma$\,Cas: 38\,km/s, ][]{2018A&A...618A.110B}
the domain of single run-away stars \citep{2019A&A...624A..66R}.

The environments of the other $\gamma$\,Cas stars in
Table\,\ref{gcasdat} were also inspected in the WISE 24\,$\mu$m atlas.
However, no convincing association of any of these stars with a nebula
was found.  In most cases, the most likely explanation is the much
larger distance implied by the much lower optical brightness.  The
field around BZ\,Cru has a very patchy background, with no structure
centered on the star standing out.

\vspace*{4mm}
In summary, there is only mild dynamic or kinematic support of the
neutron-star hypothesis for the companions to $\gamma$\,Cas and
$\pi$\,Aqr.  This makes it useful to study in more depth the role of
binarity at large in the genesis of Be stars from an observational 
(Sect.\,\ref{binobs}) as well theoretical (Sect.\,\ref{binmods}) perspective.

\section{Observations of binary Be stars}
\label{binobs}
It is not known whether all $\gamma$ Cas stars are binaries.  In view of the strongly rotationally
broadened spectral lines of Be stars and the large mass and (optical) luminosity
difference between early-type B stars and highly evolved companion
stars, attempts to prove definitively that a given Be star does not
have such a companion appear illusionary.  More quantitative statistical 
constraints, especially for less evolved systems, may result from possible discoveries 
of eclipsing systems by large-scale 
photometric monitoring surveys such as OGLE \citep{2005AcA....55..331S} or with TESS \citep{2016SPIE.9904E..2BR}.   
In any event, the assumption 
of a binary nature of $\gamma$\,Cas stars is not currently in obvious 
conflict with the available observational evidence.  

Early suggestions for a possible binary origin of Be stars were made by \citet{1975BAICz..26...65K} and \citet{1991A&A...241..419P}, triggering various observational searches.  The former work assumed that the Be disks are accretion disks.  However, owing to the lack of accreting classical Be stars, the observational support is at best weak.  Since a decretion disk can only be observed after the mass transfer, Be stars formed by mass transfer should have stripped companions  that cannot fill their Roche lobes, or their compact remnants.  

Sometimes it is even asked whether all Be stars have highly evolved companions \citep[e.g.,][]{2017ApJ...843...60W}, making their Be-typical rapid rotation the result of mass transfer from their progenitors (cf.\ Introduction).  Certainly, the scarcity of Be stars with main-sequence companions shows that, if a Be star is double, its companion very probably is highly evolved.  From a very elaborate study based on the comparison of kinematic data from Gaia for a large sample of Be stars and detailed modelling, \citet{2018MNRAS.477.5261B} infer that the 13.1\% fraction of runaway stars found by them is probably sufficient to conclude that all Be stars are post-mass-transfer binaries.  However, the apparent preference for lower-mass and highly evolved companions may bias the result if Be stars with relatively close and/or more massive companions cannot maintain a major stable disk.

In analyses of observations of individual Be stars, neutron-star, WD and sdO companions have up to now been considered almost exclusively.  The results are briefly summarised in the next three subsections. This overview may soon require 
completion for helium stars (see Sect.\,\ref{binmods}).

\subsection{Highly evolved companions}
\subsubsection{White dwarfs}
\label{WDcomp}
The first to propose that the remains of the mass donors in Be-star-forming binaries are WDs were \citet{1989A&A...220L...1W} and \citet{1991A&A...241..419P}.  Theoretical estimates of the fraction of Be stars with a WD companion reach at least 70\% \citep{2001A&A...367..848R}.  Several surveys have been conducted but no positive detection was made \citep{1992A&A...265L..41M}, with some authors considering $\gamma$\,Cas as the best candidate.  
Perhaps, a formal non-Be star, namely Regulus, currently comes closest to such systems, 
considering the late spectral subtype (B7V) and the intermittency of Be phases 
especially among late-type Be stars.  Regulus rotates about 86\% critically \citep{2005ApJ...628..439M} 
and has a WD companion \citep{2008ApJ...682L.117G}.  \citet{2009ApJ...698..666R} trace out the past and future evolution 
of this system and find that the B star may evolve into an sdB star.
In addition, \citet{2018ApJ...862..167C} recently identified some supersoft X-ray sources with Be stars in the Magellanic Clouds.  These sources are often intermittent and may be massive WDs occasionally igniting accreted matter, for example from a Be disk.  Apparently, unlike in BeXRBs, the
release of gravitational energy does not play a major role in such systems.

\subsection{sdO stars}
\label{sdOcomp}
Two of the first Be stars initially suspected and later demonstrated
to be orbited by a low-mass star strongly interacting with the Be disk
were HR\,2142 \citep{2016ApJ...828...47P} and $\phi$\,Per
\citep{2015A&A...577A..51M}.  In UV spectra (mostly from {\it IUE})
with sufficient orbital phase coverage, spectral lines can be clearly
seen with a much larger velocity swing than that of the B-type primary
\citep{1995ApJ...448..878T}.  Numerous narrow Fe\,IV, V, and VI lines
as well as the He\,II $\lambda$1640 line convincingly show the
similarity to spectra of sdO stars \citep{1998ApJ...493..440G}.  A
more indirect indicator of a hot companion can be a hot spot in the
disk where helium emission lines form.  Periodic shifts in radial
velocity trace the secondary's orbit \citep{2004A&A...427..307R}.

In stars with only a few scattered UV spectra, cross-correlations of
the observations with model spectra have been used to identify
additional sdO companions and candidates \citep{2018ApJ...853..156W}.
As explained in Sects.\,\ref{gCas} and \ref{piAqr}, this method is not
very effective for broad-lined early-type Be stars, i.e., many
$\gamma$\,Cas stars.  The relatively low detection rate is probably
also due to the low S/N ratio of {\it IUE} spectra.  The total number
of Be stars with a detected or likely sdO companion is about 15
\citep{2018ApJ...853..156W}.

\subsection{Neutron stars and black holes}
Systems with neutron-star and black-hole companions 
\citep[currently, only one Be system with a black hole seems to be known][]
{2014Natur.505..378C}, i.e.\, BeXRBs are omitted from the discussion because, as outlined above, the X-ray properties of $\gamma$\,Cas stars seem incompatible with those of BeXRBs and there is no convincing evidence that the companions of $\gamma$\,Cas and $\pi$\,Aqr are neutron stars or even black holes.  
However, the immediate progenitors of BeXRBs, namely Be stars with a helium-star companion have not yet been placed into a close perspective with the formation of Be stars; this will be done in Sect.\,\ref{binmods}.

\section{Binary stellar evolution models}
\label{binmods}
\begin{table*}
  \caption{Key data of selected massive binary evolution models from
    \citet{1999A&A...350..148W} and \citet{2001A&A...369..939W}.  Besides the initial binary parameters, i.e.,
    the initial masses of the mass donor ($M_{\rm 1,i}$) and the mass gainer ($M_{\rm 2,i}$), and the initial orbital period $P_{\rm orb,i}$,
    we give parameters of the binary and its component stars at the time where the
    mass donor has a central helium mass fraction of 0.8 during core helium burning, i.e.,
    the orbital period $P_{\rm He+OB}$, both stellar masses during that stage, the corresponding luminosities
    and effective temperatures, and the expected stellar wind mass loss rate, velocity and mechanical wind energy production rate
    according to \citet{2017A&A...607L...8V}.}
\label{Lmodels}
\centering\begin{tabular}{c c c c c c c c c c c c c c c c c}
\hline\hline
No. & $M_{\rm 1,i}$ & $M_{\rm 2,i}$ & $P_{\rm orb,i}$ & $P_{\rm He+OB}$ & $M_{\rm He}$  & $M_{\rm OB}$ & $L_{\rm He}$  & $L_{\rm OB}$    & $T_{\rm He}$ & $T_{\rm OB}$ & log $\dot M_{\rm He}$ & $\varv_{\rm esc,He}$ & $L_{\rm wind,He}$ \\
 ~ & M$_\odot$ & M$_\odot$ &     d   &   d    & M$_\odot$ & M$_\odot$ & 10$^3$ L$_\odot$ & 10$^3$ L$_\odot$ &    kK      &    kK  & M$_\odot$/yr      & km/s       & L$_\odot$  \\
\hline
1 &  12      &  8       &   2        &    189       &  1.1     &   18.5   &   0.436  &  69 &    48      &    32     &   -9.71          &  1170      &   0.40   \\
2 &  10      &  8       &   3        &     61       &  1.8     &   16.1   &   2.7    &  29 &    57      &    32     &   -8.63          &  1130      &   2.25   \\
3 &  12      &  8       &   6        &     72       &  2.4     &   17.5   &   6.2    &  35 &    65      &    33     &   -8.14          &  1200      &   7.74   \\
4 &  16      & 13       &   3        &     64       &  2.6     &   25.3   &   6.9    & 120 &    74      &    35     &   -8.08          &  1390      &  12.00   \\
5 &  16      & 15       &   9        &    107       &  3.6     &   26.7   &  16.6    & 126 &    80      &    38     &   -7.56          &  1420      &  41.20   \\
6 &  25      & 19       &   4        &     30       &  5.3     &   35.9   &  35.0    & 238 &    98      &    43     &   -7.12          &  1750      & 180.10   \\
\hline
\end{tabular}
\end{table*}

Since stars, during their evolution, tend to increase their radii by
large factors, most close binary systems will experience transfer of mass between the two stars.
For the closest binaries, i.e.\ for orbital periods typically below $\sim10\,$d, mass transfer starts
while both stars undergo core hydrogen burning \citep[Case A;][] {1994A&A...288..475P, 1994A&A...290..119P, 
2001A&A...369..939W}. In this case, the mass transfer is divided into three distinct phases:
a thermal-timescale mass transfer (fast Case\,A), which is succeeded by a nuclear-timescale mass transfer phase during which the mass
ratio is inverted (slow Case\,A or Algol phase), followed by another thermal-timescale mass-transfer once the donor star 
ends core hydrogen burning (Case AB). In wider binary systems, the post-main sequence expansion of the
initially more massive star leads to thermal-timescale mass transfer, while the companion is generally still burning hydrogen
(Case\,B). In both cases, the mass transfer may become unstable, with the likely consequence of a merger of
both stars \citep{2014ApJ...782....7D}. However, if a merger is avoided, the mass donor -- the initially more massive star -- 
loses almost its entire hydrogen-rich envelop due to mass transfer, while the mass gainer is accreting all or only part of it.
The ratio of the number of mergers and the number of stable mass-transfer systems, and the mass-transfer efficiency, are uncertain
\citep{2012ARA&A..50..107L}.

\citet{1963PASP...75..207S} and \citet{1966ARA&A...4...35H} realised that the accretion of mass from a
companion star can lead to an increase of the star's specific angular momentum, with the
consequence that mass gainers may spin supersynchronously w.r.t. the orbital rotation.
This effect is observationally well documented for massive Algol systems \citep[e.g.,][Mahy et al., submitted]{2015A&A...582A..73H}, where spin-up to critical rotation is avoided through tidal spin-orbit coupling
\citep{2013ApJ...764..166D}.

Only after Case\,AB or Case\,B mass transfer, if the mass transfer is not too inefficient, does one expect
the spin-up process to drive the mass gainer towards critical rotation, since the orbits become wide enough to
render tides negligible. It was shown analytically by \citet{1981A&A...102...17P},
and later through detailed models by \citet{2005A&A...435.1013P}, that a mass increase by only 10\%
can be sufficient to spin up a star to its critical rotation. The problem with this situation is that
after Case\,AB or Case\,B mass transfer, the envelope mass of the donor has become very small such that the
donor is hotter than a main sequence star, and thus remains very faint in optical light. Furthermore, its remaining
lifetime is much shorter than that of its spun-up companion. This means, it will rapidly evolve into a compact object,
which, in case a neutron star or black hole is formed, may lead to the disruption of the binary by the supernova explosion. As a consequence,
most post Case\,AB or Case\,B systems may not be recognised as such \citep{2014ApJ...782....7D}. 

In the following, it will be assumed that the mass gainers of Case\,AB or Case\,B are in fact spun up such that they
appear as Oe/Be stars after the mass transfer. This idea is, of course, strongly supported by the large number of
classical BeXRBs, which are explained as such post-Case\,AB or Case\,B binaries where the donor star evolved into 
a neutron star without breaking up the binary \citep{2006csxs.book..623T}. In these systems, the nature of the companion
is revealed by the copious X-ray emission which is produced when the neutron star crosses  
or approaches the Be disk in its tilted and/or elliptical orbit,
which leads to mass accretion onto the neutron star. As seen in Sect.\,\ref{binobs},
there is also a smaller number of Be\,stars with known or suspected BH, WD or sdO companions, 
which all fit into the post-mass transfer scenario.

It is worth pointing out that rotating, non-magnetic stars can spin down due to stellar-wind mass loss
\citep{1998A&A...329..551L}. For the most massive main sequence stars, which lose a significant fraction
of their initial mass through a wind, this process may be efficient, and observational evidence for this exists
in Galactic O\,stars \citep{2018A&A...613A..12M}. It explains also the fast but sub-critical
rotation of the O\,stars in Galactic WR+O-star binaries \citep{2018A&A...615A..65V},
in which the WR star was likely the mass donor in a mass transfer process \citep{2005A&A...435.1013P}. However, 
$\gamma$\,Cas stars are Be/Oe stars which do not spin down. This is consistent with
the expectation that the main sequence mass loss in Galactic stars is below 10\% for stars with an
initial mass below 28\,M$_{\odot}$ \citep{2011A&A...530A.115B, 2012ARA&A..50..107L}.


\begin{figure*}
\includegraphics[width=13.9cm,angle=-90]{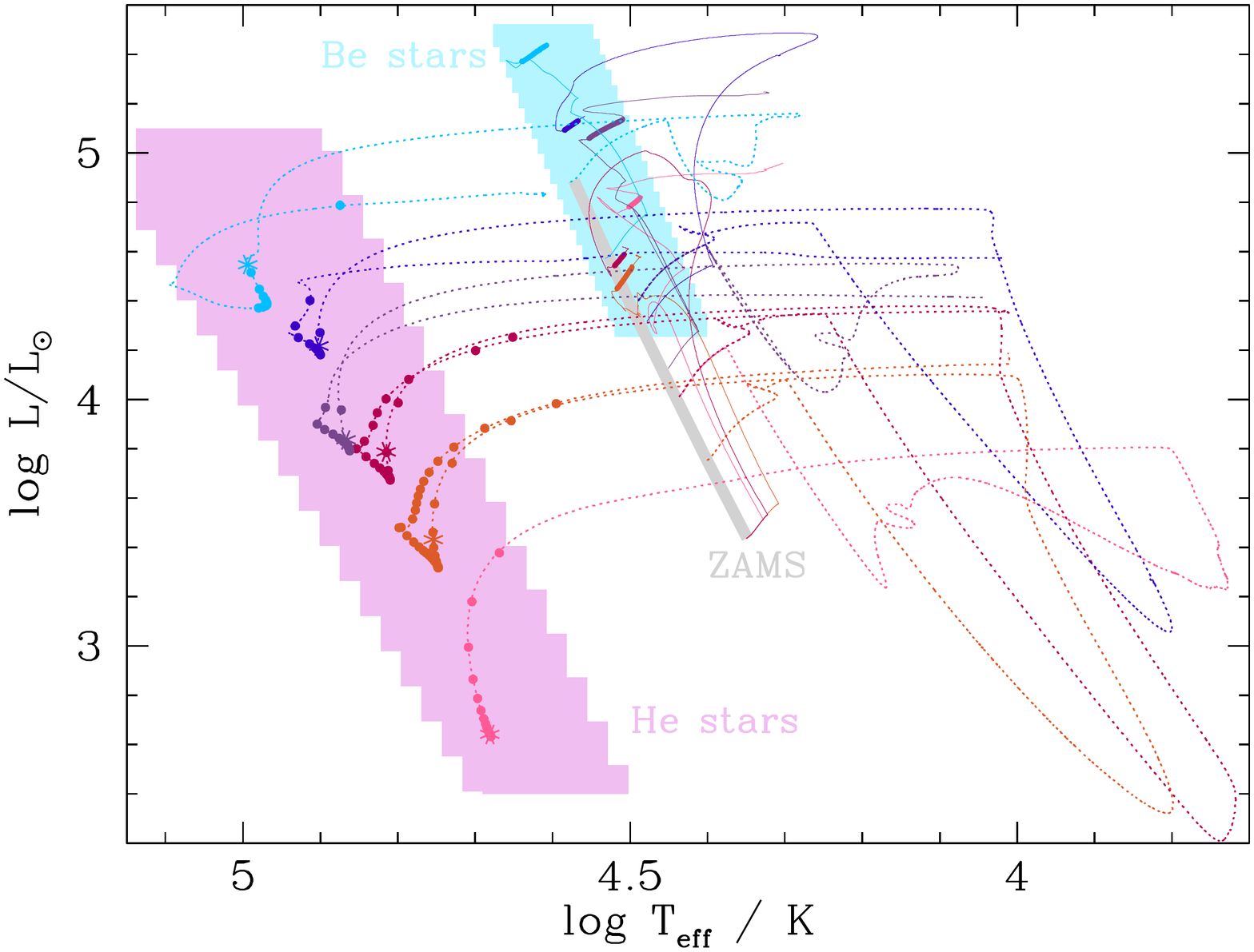}
\centering
\caption{Evolutionary tracks of both, the mass donors (dotted lines) and 
the mass gainers (solid lines) of the analysed six
binary models in the Hertzsprung-Russell diagram. 
Pairs of tracks with the same colour belong to the same binary system.
The thick gray line marks the zero-age main sequence for the initial mass range of
our models (i.e., from 8 to 25\,M$_{\odot}$).
The thick solid drawn parts of the mass gainers' tracks marks the
phase during which the companion is a He star (starting from a
core helium mass fraction of $Y_{\rm c}=0.95$ during core helium burning).
The corresponding area in the HR diagram is coloured light blue and labelled 
``Be stars''. On the tracks of the mass donors, dots are placed every $10^5\,$yr during core helium burning,
and a star symbol marks the time at which their core helium mass fraction
is $Y_{\rm c}=0.80$. The tracks end during the phase of shell helium burning
with a small remaining lifetime of the He stars, except for the System No.\,1, 
which ends at $Y_{\rm c}=0.7$. The area in the HR diagram in which the He star models
spend most of their lifetime is coloured in pink and labelled ``He stars''.
The tracks correspond to the binary models Nos.\,1 to 6
(Tab.\,2) in the order of increasing helium-star luminosity (as marked by the star symbols).}
  \label{hrdx}
\end{figure*}

\subsection{The case for $\gamma$\,Cas stars as Be + helium-star binaries (BeHeBs)}

According to the above considerations, an Oe/Be star, when it is formed as such 
in a binary system, has a helium star companion. The corresponding Be$+$helium-star binaries will
below be called BeHeBs for short. While the helium star evolves faster
than the Be\,star, the lifetime of this BeHeB stage -- the helium burning timescale 
of the helium star -- is long enough to expect that some of the observed Be binaries
are in this stage (cf.\, Sect.\,\ref{number}). 

The following will discuss
the hypothesis that $\gamma$\,Cas stars are BeHeBs, based on binary evolution models 
computed by \citet{1999A&A...350..148W} and \citet{2001A&A...369..939W}. Whereas these
models do not include rotation, they assume conservative mass transfer, which implies
that the mass increase is sufficient to spin up the mass gainer to critical rotation.
Table\,\ref{Lmodels} gives an overview of the initial parameters of these models, and those during the BeHeB
stage. These were chosen such that the masses of the formed helium stars
($1.1-5.3\,$M$_{\odot}$) cover the plausible mass range of such objects in $\gamma$\,Cas stars.
That is, more massive helium stars would likely form optically thick winds, which would make them
easily identifiable as Wolf-Rayet stars \citep{1989A&A...210...93L}. And helium stars significantly below $1$M$_{\odot}$
even require progenitors of so low initial mass that the mass gainer could not evolve into a B star of the
earliest spectral type. 

Figure\,\ref{hrdx} gives an overview of the evolution of both components of the binary models in the Hertzsprung-Russell diagram.
The tracks of the pairs of stars start on the zero-age main sequence. Whereas otherwise these
evolutionary tracks show the typical pattern of Case\,A and\,B binary models \citep[cf.\,][]{2001A&A...369..939W},
the thick-drawn part of the lines focuses on the BeHeB stage, i.e., on the time period during which the mass donor
evolves through core helium burning. 

As the mass gainers -- the presumed later Be stars -- hardly evolve during
this time, the thick-drawn stretch of their evolutionary tracks is very short. 
For the mass donors, however, there is significant evolution. In any case, it is important to realise
that the donors move fast along the horizontal parts of the evolutionary tracks.
The thick dots on their tracks mark a central helium mass fraction of $Y_{\rm c}=0.8$, and core helium
exhaustion is signified by the end of the thick-drawn part of the track. Therefore,
the time-averaged properties of the helium stars are well represented by their properties
at $Y_{\rm c}=0$. 

It follows from Fig.\,\ref{hrdx} that the helium-star
secondaries will be difficult to observe, in the optical and at longer wavelengths, next to the much brighter Be
star \citep{2018A&A...615A..78G}. However, with luminosities of 500 to 50,000\,L$_{\odot}$, helium
stars are still luminous stars, and as such they are expected to emit a
radiation-driven wind.  Observational evidence for this is found in the UV
spectra of the rare so-called extreme helium stars \citep{2010MNRAS.404.1698J}. 
While helium-star wind mass loss based on \citet{1982A&A...116..273H} is included in the 
presented binary evolution models, the present study uses the recent theoretical
mass-loss rates by \citet{2017A&A...607L...8V}, which reproduce the
empirical rates of Hamann et al.\ reasonably well, but also provide a
smooth transition to the mass-loss properties of the more massive
Wolf-Rayet stars. As the total amount of mass lost during core helium
burning is mostly very small, this does not introduce any significant inconsistency.

The models provide guidance in answering the question whether the
presence of a wind emanating from the helium star could give rise to
an observable X-ray signal in BeHeBs. 
As the helium stars are compact, and their winds fast, the models lead
to the expectation of X-ray emission from two potential interaction
regions. The first candidate zone is where the wind of the He star
encounters the disk of the Be star, and the second one resides where
the He-star wind meets the -- also present -- ordinary
radiation-driven wind of the Be star. The following subsections examine
these two cases.

\subsubsection{Interaction between He-star wind and Be disk}
\label{winddisk}

If the He star had no wind, the Be disk might well extend to the He
star companion or even engulf it.  This is so by analogy to the BeXRBs,
where a neutron star, i.e., the descendant of a helium
star in a BeHeB, emits X-rays when it crosses the equatorial plane of
the Be star. Since He stars possess a
strong wind, they will blow a cavity into the Be disk, whose
size may be determined by the balance of the wind ram pressure and the
thermal and turbulent pressure of the gas in the Be disk. The cavity may be
elongated in the direction of the orbital motion, and its vertical
size will depend on the thickness and vertical structure of the Be
disk.  Truncation by the companion of the disk \citep{2018MNRAS.473.3039P} 
could lead to still other geometries.  Some fraction of the He-star
wind could escape without interacting with the disk.  

In any case, the interaction shock front will likely have a complex
three-dimensional structure, and may develop turbulence and magnetic
fields, which would all affect the emission of energetic photons. In a
first simple step, the next paragraphs attempt to derive upper limits
on the X-ray luminosity and the photon temperature from predictions of
stellar-evolution and radiation-driven-wind physics.

\begin{figure}
\includegraphics[width=6.9cm,angle=-90]{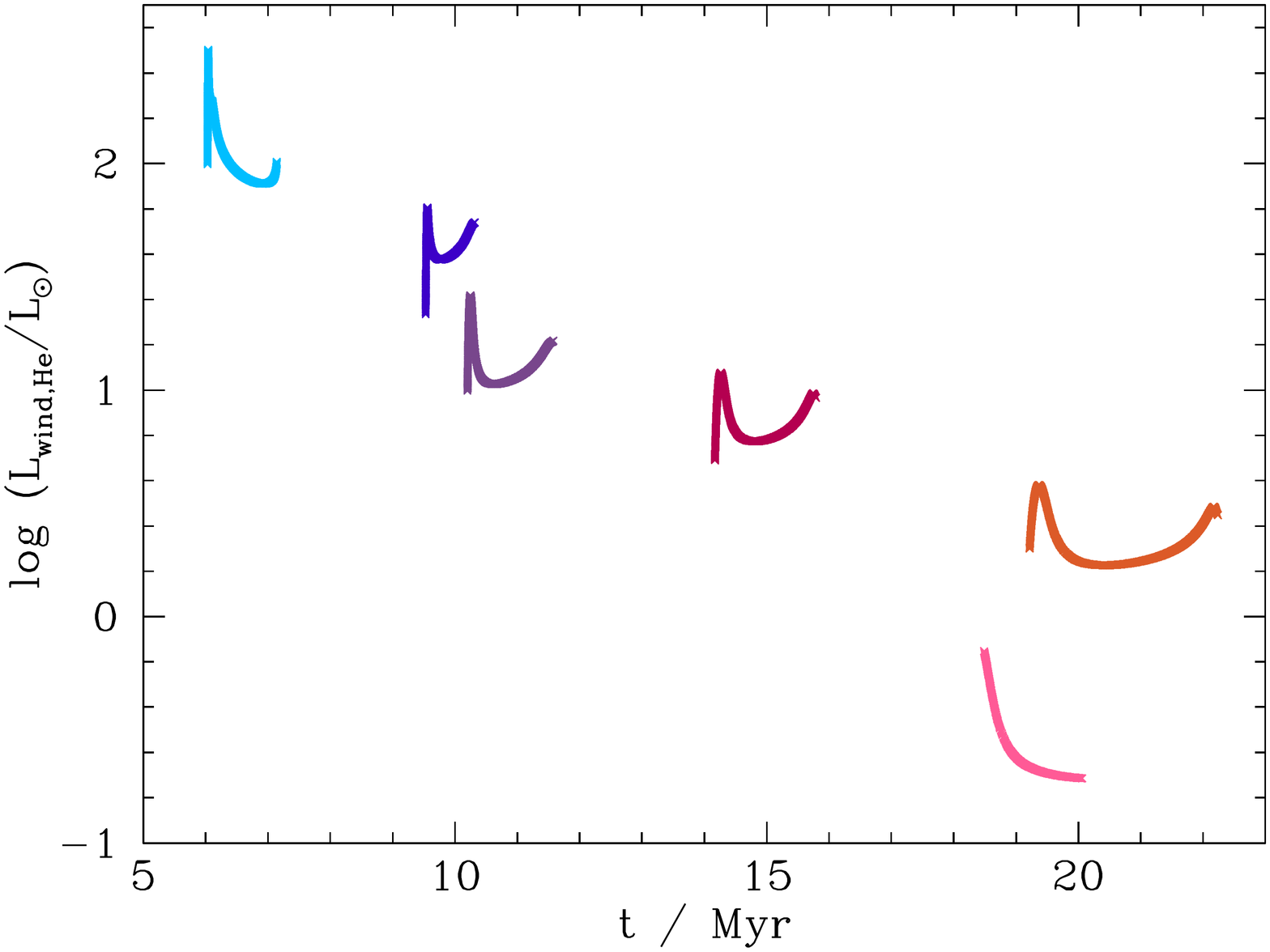}
  \caption{Time evolution of the mechanical luminosity of the donor
    star's wind during its helium-star stage, for the six binary models
    during core helium burning of the donor star. The colour coding is
    the same as in Fig.\,\ref{hrdx}, and the tracks belong to binary models Nos.\ 1 to 6
    (Table\,\ref{Lmodels}) in increasing order of their wind luminosity.} 
\label{lx}
\end{figure}

Figure\,\ref{lx} illustrates the time dependence of the helium star's
mechanical wind luminosity $L_{\rm wind, He} = {1\over 2} \dot M_{\rm He} \varv_{\rm wind, He}^2$ 
for the six evolutionary models in Table\,\ref{Lmodels}. Here, $\dot M_{\rm He}$ is the mass loss rate predicted by
\citet{2017A&A...607L...8V}, and $\varv_{\rm wind, He}$ is the terminal wind
velocity, for which Vink showed that it exceeds the escape speed of the
helium stars by about a factor of three. It is, therefore, assumed here that
$\varv_{\rm wind, He}=3 \sqrt{2 G M_{\rm He} /R_{\rm He}}$.

Figure\,\ref{lx} also provides an upper limit to the X-ray luminosity produced by the
wind-disk interaction because only a fraction of the kinetic energy can be 
converted to X-rays.  As the figure shows the wind kinetic-energy fluxes are of the
order of a few hundred L$_{\odot}$ for the massive helium stars
($M \simeq 5$\,M$_{\odot}$) down to fractions of L$_{\odot}$ at lower
masses ($M \simeq1$\,M$_{\odot}$).  These numbers should only be
taken as order-of-magnitude indicators since, in his pioneering study,
\citet{2017A&A...607L...8V} adopted a fixed effective temperature of
$50\,000\,$K ($\log T_{\rm eff} \simeq 4.7$; cf.\ Fig.\,\ref{hrdx})
while the temperature dependence of these winds is not yet well
understood. Clearly, even lower wind luminosities will occur in systems 
with masses below the range considered here.  However, as potentially observable effects will
become correspondingly weaker, they are not considered here.  More
massive systems, on the other hand, might contain O stars whose strong
winds would -- at least at Galactic metallicities -- spin down the
stars, such that they would not be Oe/Be stars for long. 
Potentially, they would also destroy any circumstellar disk.

As for the X-ray luminosity, the given models only place an upper limit 
on the temperature of the hot gas which is produced by the
shock front where the He-star wind hits the Be disk. 
With escape speeds of the helium stars in the range
1100-1800\,km/s, the terminal wind speeds of the He star are of the order of 
3000-5000\,km/s (see above). For an adiabatic shock, these numbers 
translate to temperatures of about $5 \cdot 10^8\,K$ to $15 \cdot 10^8\,K$, or 50 to 130\,keV,
assuming $T=m_{\rm p} \varv_{\rm wind, He}^2/(2k)$, where $m_{\rm p}$ is the mass of the proton.  
On the other hand, at sufficiently high
densities, the shock may be non-adiabatic so that the achieved temperature can
be much smaller.  Hydrodynamic instabilities, clumping, or entrainment
of cold gas may as well lead to smaller temperatures.  
However, this is challenging to estimate quantitatively, 
and beyond the scope of the present work.

\subsubsection{Interaction between He-star wind and Be-star wind }
\label{windwind}

As mentioned above, a fraction $f_1 < 1$ of the He-star wind may be
able to escape without interacting with the Be disk. Part of this
matter will, however, collide with the ordinary Be-star
wind, so that only a fraction $f_1 * f_2$ of the He-star wind leaves
the system without any interaction at all. Here, $f_2 <1$ designates
the fraction of the He-star wind not hitting the Be disk that also
escapes collision with the Be-star wind.

\begin{figure}
\includegraphics[width=6.9cm,angle=-90]{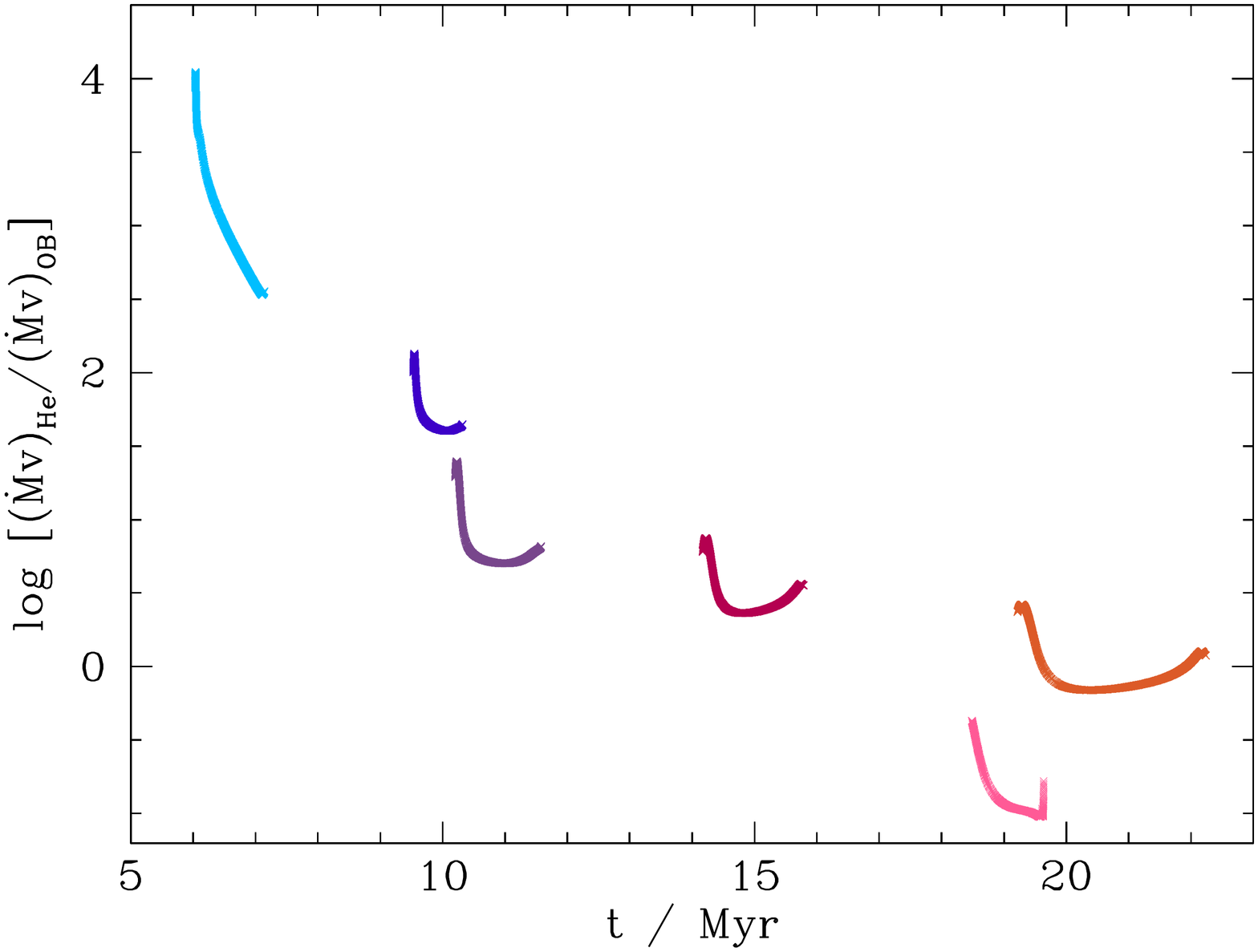}
\caption{Time evolution of the ratio $\eta$ of the wind momentum of the donor
  star to that of the mass gainer for the six binary models (see
  Table\,\ref{Lmodels}) during core helium burning of the donor
  star. The colour coding is the same as in Fig.\,\ref{hrdx}.}
\label{p12-t}
\end{figure}

A key parameter determining the interaction fraction and also the
X-ray production efficiency of colliding wind systems is the wind
momentum ratio
$\eta = \dot M_{\rm He} \varv_{\rm He} / \dot M_{\rm OB}
\varv_{\rm OB}$, where $\dot M$ and $\varv$ denote the mass-loss
rates and terminal wind velocities of both stars, respectively.  The interaction
fraction and the X-ray production efficiency are largest for
$\eta = 1$ \citep{2018MNRAS.477.5640P}.  Figure\,\ref{p12-t} illustrates the time dependence of
$\eta$ for the selected binary-model sequences during the stage of
core He-burning of the helium star.  The underlying mass-loss
rates and terminal wind velocities are those proposed by
\citet{2014A&A...564A..70K} for ordinary B main-sequence stars. 
As the terminal wind speeds of Krti{\v c}ka's wind models
are roughly three times the corresponding escape speed from the star,
the escape speed of the models in Table\,\ref{Lmodels} was multiplied by 
a factor of three to
compute their terminal wind speeds.  This 
neglects the possibility that in very close systems, or for values of $\eta$
far from unity, one or both winds might not quite have attained their
terminal speeds when reaching the interaction point.

Figure \ref{p12-t} demonstrates that, in the considered binary models,
quite diverse situations may prevail. In some systems (typically the
more massive ones) the He-star wind momentum is larger by more than
an order of magnitude, in some other systems (typically the less
massive ones) the B-star wind is stronger by a similar factor, and in
still others the wind momentum ratio is close to unity. This occurs
because both the wind velocities and the mass-loss rates of He- and
B-stars are not too different.

Similar to the wind-disk interaction, it is difficult to provide
firm predictions for the X-ray emission produced by the wind-wind
interaction. The upper limits to the X-ray flux and plasma temperature
are similar to those of the wind-disk interaction, since the wind
velocities and mass-loss rates are also similar. On the other hand, 
the conditions in the wind-wind interaction region will be different
from those in the wind-disk case, since, e.g., the matter density 
in the Be disk will be larger than that in the Be star wind. It is therefore
possible that X-ray emission will be composed of more than one discrete component.

Observations of
colliding-wind binaries show that the X-ray flux in massive
He-star+-O-star binaries can reach about 100$\,$L$_{\odot}$ and
temperatures up to 100\,MK \citep{2012ASPC..465..301G}. The mass-loss
rates in these systems are several orders of magnitude above those in
BeHeBs, but the wind velocities are
comparable. \citet{2018MNRAS.477.5640P} find from hydrodynamical
simulations that the expected X-ray emission decreases roughly
linearly with the weaker-to-stronger wind-momentum ratio.  Moreover,
in BeHeBs, the wind-wind interaction is restricted to higher
latitudes, as the equatorial regime is blocked by the Be
disk. Nevertheless, while there are several factors which may reduce
the X-ray emission, a detectable X-ray flux from the wind-wind
interaction is not excluded, especially not for BeHeBs with a wind
momentum ratio near unity, which is achieved by the majority of the 
models considered (Table\,\ref{Lmodels}).

\subsubsection{The expected number of BeHeBs}
\label{number}
%

The BeHeB phase is a short intermediate evolutionary phase of massive binary
systems, for which direct observational evidence is still
lacking.  This phase is defined by the core helium-burning stage of
the mass donor. It is often disregarded in comparison with observations,
because its duration is mostly shorter than that of the foregoing Algol phase
(if any), but also shorter than the subsequent BeXRB or Be$+$WD phase.

The binary evolution models of \citet{2001A&A...369..939W} can give an
estimate of the number of BeHeBs relative to BeXRB or Be$+$WD systems.
The stripped core helium-burning companions to B-star mass
gainers are very hot ($T_{\rm eff} \simgr 50 kK$) and sub-luminous
(cf.\ Fig.\,\ref{hrdx}), leading to no realistically observable signal in the
optical regime which is dominated by the B star.  The lifetime of the
faint, hot helium star is the nuclear timescale of core helium
burning, which is of the order a few Myr for helium stars in the mass
range $1.6-5\,$M$_{\odot}$ \citep{2019ApJ...878...49W}.
Compared to the hydrogen-burning lifetimes
of the rejuvenated B-star mass gainers of 10 to 30\,Myr, this is about
10\% or less.  Therefore, among the Be stars that have emerged from
this binary evolution channel, a comparable fraction, i.e., up to 10\%, could have a
helium-star companion.

For an accurate prediction of the number of $\gamma$\,Cas binaries
expected from the ansatz pursued above a population-synthesis study
will probably be required.  Clearly, the estimate of $\sim$10\% for
the number ratio of BeHeBs to Be binaries with compact companions (BeXRBs and
Be$+$WD systems) can serve as an upper limit.  However,
direct empirical comparisons will suffer from a strong observational
bias.  $\gamma$\,Cas binaries would be identified on account of their
X-ray properties while only $\gamma$\,Cas binaries with sufficiently
massive helium-star companions are predicted to have detectable X-ray
fluxes.  The latter subpopulation may roughly consist of those systems
in which the helium stars end their evolution as neutron stars. The
expected fraction of observed $\gamma$\,Cas binaries would be 10\% of
the BeXRBs, multiplied by the luminosity bias factor, and divided by
the break-up fraction, $f_{\rm breakup}$ of Be binaries at neutron-star formation in a
supernova explosion.  Both factors are quite uncertain, but a fraction
of about 10\% of all BeXRBs (i.e., of all progenitor systems not disrupted by 
a supernova explosion) does not seem impossible.

\subsection{Comparison with observations}
                                 
As seen above, some fraction of the Be-star binaries (the BeHeBs) are
expected to contain a core helium-burning star. The He star is not
likely to be readily observable as it is bolometrically much dimmer
than the B star.  Because the He star is much hotter than the Be
star, the contrast problem is lowest in the UV.  The previous section
considered corresponding binary-evolution and stellar-wind models,
with the idea in mind that the presence of a helium star may give rise
to observable X-ray emission. The following discusses to what extent
the $\gamma$\,Cas stars and their peculiar X-ray properties (cf.\
Sect.\,\ref{synopsis}) might correspond to the BeHeBs. To this effect,
the recent compilation of $\gamma$\,Cas stars by
\citet{2018A&A...619A.148N} is used, from which the quantities in
Table\,\ref{gcasdat} were drawn.

Firstly, it should be noted that, in BeHeBs, there may be other sources of X-rays
than those which are induced by the fast wind of the helium star.
In particular, the helium star itself can be so hot that it emits
X-rays. E.g., for $T_{\rm eff}=100\,000\,$K, the Planck function peaks
at 0.3 keV. As this is the hottest temperature expected for
BeHeBs, it follows that only very soft X-rays can be produced in this way.
This holds similarly for the thermal emission of hot pre-white dwarfs,
as well as for accreting white dwarfs \citep[cf.\,][]{2018ApJ...862..167C}.
As the X-rays measured in $\gamma$\,Cas stars are much harder 
(cf.\ Sect.\,\ref{gCstars}), they are unlikely to be produced in stellar
photospheres.  

\begin{figure}
\includegraphics[width=6.9cm,angle=-90]{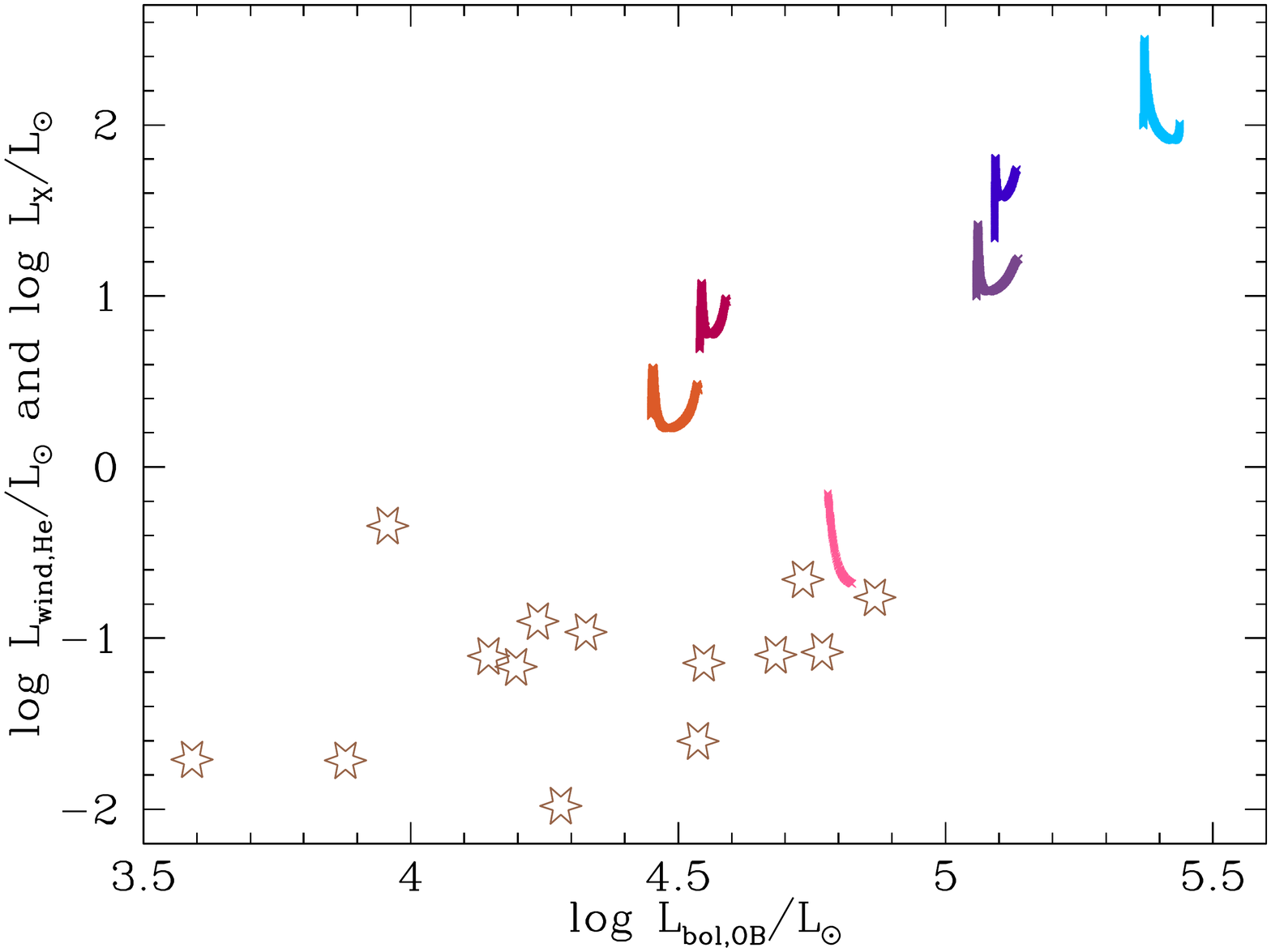}
\caption{Mechanical luminosities of the donor star winds versus the
  mass gainer's bolometric luminosity for the six binary models (cf.\
  Table\,\ref{Lmodels}), during core helium of the donor star. The colour
  coding is the same as in Fig.\,\ref{hrdx}, and the tracks belong to binary models Nos.\,1 to 6
    (Table\,\ref{Lmodels}) in increasing order of their wind luminosity.  Also plotted are the
  X-ray luminosities of the $\gamma$\,Cas stars 
  versus their bolometric luminosities, according to
  \citet[][see also Table\,\ref{gcasdat}]{2018A&A...619A.148N}.}
\label{lx-lbol}
\end{figure}


Secondly, in Fig.\,\ref{lx-lbol}, the mechanical luminosities of the
helium-star winds as a function of the OB star's bolometric luminosity
for the six model systems (Table\,\ref{Lmodels}) are plotted together with the observed X-ray
luminosities of the $\gamma$\,Cas stars and their respective
bolometric luminosities.  Here, it should be stressed that both, predicted
(cf., Sects.\,\ref{winddisk} and\,\ref{windwind}) and observed quantities 
(Table\,\ref{gcasdat}) have large uncertainties.
In particular the bolometric luminosities of the Be\,stars could
be wrong a factor of\,$2-3$, since their luminosity classes are
only adopted, individual extinction corrections have not been determined,
and the rapid rotation of the Be stars leads to an anisotropy
of the photon emission \citep{1924MNRAS..84..665V} which is unaccounted.

There is some overlap in the areas
populated by models and observations in Fig.\,\ref{lx-lbol}.
Also, the bolometric luminosities of both samples span about
a factor of 20, and the ordinate range of models and observations span
$2-3$\,dex. However, there are considerable offsets in both coordinates 
between the two datasets, namely by about a factor of 3 in $L_{\rm bol, OB}$,
and a factor of 100 comparing $L_{\rm wind, He}$ and $L_{\rm X}$.

Specifically considering the ordinate of Fig.\,\ref{lx-lbol}, the difference may
arise because the conversion of mechanical wind energy to X-rays in
$\gamma$\,Cas binaries is an inefficient process. In fact, this
seems to be generally the case for colliding-wind binaries as
discussed above, in particular when the wind momentum ratio is far
from unity. The conversion efficiency in the wind-disk interaction
case is less clear owing to the lack of comparable cases in other
systems. However, we can conclude that the mechanical wind energy
of the helium stars is sufficient to account for the observed
energy in X-rays in $\gamma$\,Cas stars.
                                 
The range in OB-star bolometric luminosities (the abscissa in
Fig.\,\ref{lx-lbol}) should be more directly comparable, in spite of the caveats mentioned above.
In $L_{\rm bol, OB}$, the overlap between models and observations is larger,
but the models are generally more luminous. One reason for the
difference is the inclusion of a fairly massive model (No.\,6 in Table\ref{gcasdat}), mostly
for illustrative purposes, as this may correspond better to WR+O-star
binaries instead of BeHeBs. Certainly, at 36$\,$M$_{\odot}$ the mass
gainer in this model becomes so massive that its wind will spin it
down quickly \citep{2011A&A...530A.115B} so that its lifetime as
Oe star would be very short. However, it is also important to consider that the
models of \citet{2001A&A...369..939W} are mass conserving, which means
that the entire mass lost from the donor is accreted onto the mass
gainer.  Recent evidence shows, however, that mass transfer in massive
close binaries may well be non-conservative on average
\citep{2007A&A...467.1181D, 2012ARA&A..50..107L}.  Because
non-conservative evolution does not lead to a different evolution
for the donor stars, the He-star properties of
\citet{2001A&A...369..939W} would remain about the same in the non-conservative case. 
However, the mass gainers, which are mostly in the late O-star regime in the models
analysed above, would be significantly less massive, and thus less
luminous. That is, the tracks of the models shown in
Fig.\,\ref{lx-lbol} would move to the left at constant ordinate. In
extreme cases, the mass gainer's mass would just be about half of what
it becomes in the conservative model, thereby decreasing
$\log L/$L$_{\odot}$ by about 0.9\,dex.

It seems unlikely that models and data could be brought into agreement
by considering binary models with smaller initial masses. As seen in
Fig.\,\ref{lx-lbol}, models and data might overlap well if the
downward trend of wind luminosity with bolometric luminosity
continued. However, if the observed sample of $\gamma$\,Cas stars is
merely the peak of a distribution which extends to lower X-ray
luminosities, the known $\gamma$\,Cas stars should correspond to the
most luminous models that predict the $\gamma$\,Cas phenomenon.  This
range is obviously covered by the chosen theoretical
tracks. 

In summary, if $\gamma$\,Cas stars are binaries with core
helium-burning helium stars, Fig.\,\ref{lx-lbol} suggests that (a) the
mass transfer efficiency during the preceding mass transfer phases was
about 0.5 (since the model tracks would have to be shifted to the
left by about 0.4\,dex to match the data) and (b) about 1\% of the
wind luminosity would be converted into X-rays.

\begin{figure}
\includegraphics[width=6.9cm,angle=-90]{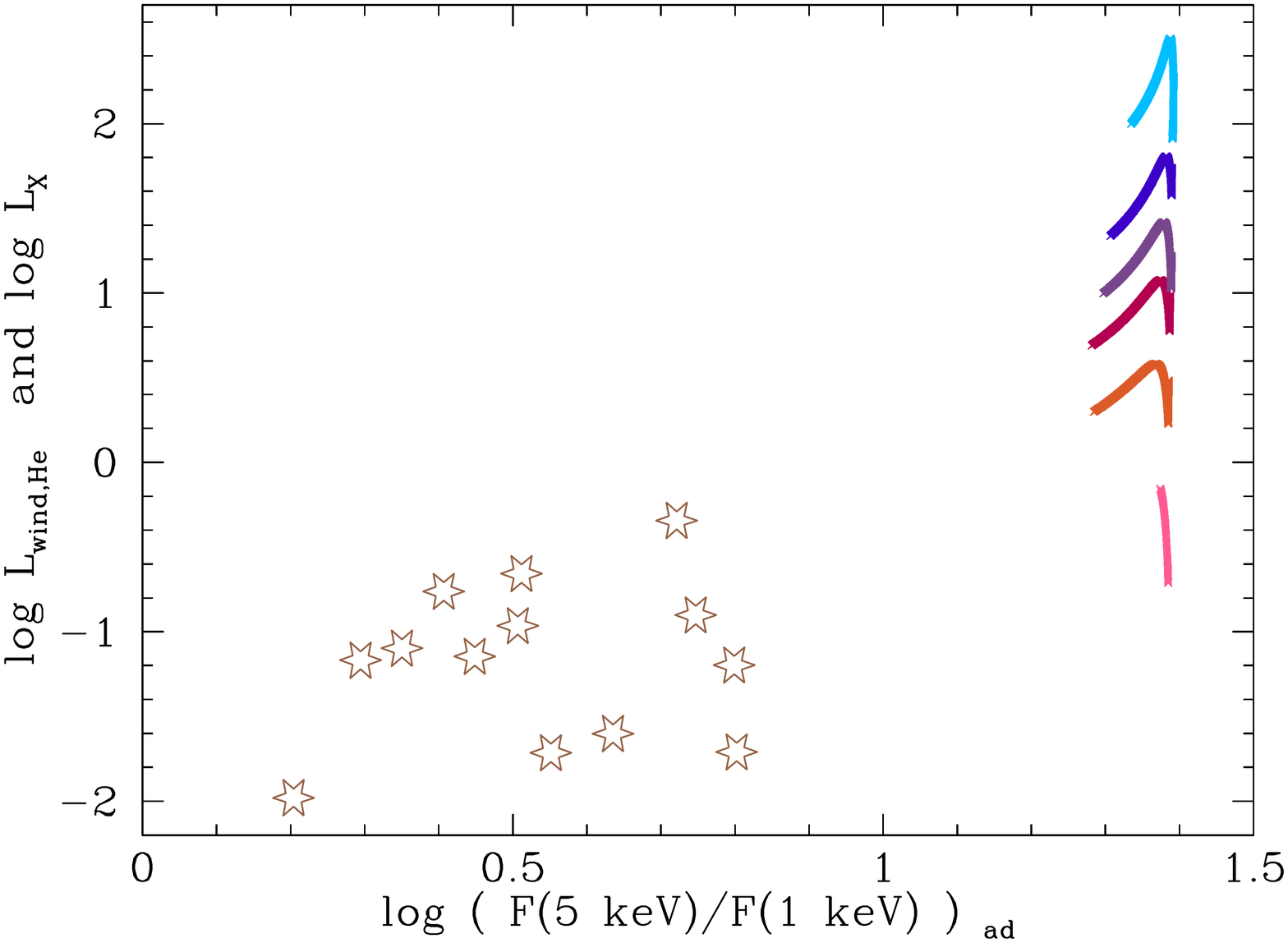}
\caption{Mechanical luminosities of the donor-star winds versus the
  5\,keV/1\,keV flux ratio assuming adiabatic shocks for the six binary
  models (cf.\ Table\,\ref{Lmodels}), during core helium burning of the donor
  star. The colour coding is the same as in Fig.\,\ref{hrdx}.  Also
  plotted (as stars) are the observed X-ray luminosities and 5\,keV/1\,keV flux
  ratios of the $\gamma$\,Cas stars versus their bolometric luminosities, according to
  \citet{2018A&A...619A.148N} (cf.\ Table\,\ref{gcasdat}).  }
\label{lxloghr}
\end{figure}

Figure\,\ref{lxloghr} compares the (mechanical) He-star wind
luminosities and the 5\,keV/1\,keV flux ratio using the black-body
approximation and the adiabatic post-shock temperature of the winds
with empirical data of $\gamma$\,Cas stars from
\citet{2018A&A...619A.148N}. Taken at face value, the X-ray
luminosities as well as the adiabatic flux ratios derived from the
models (Table\,\ref{Lmodels}) are much too high compared to the
observed X-ray luminosity and hardness ratio.  However, as discussed
above, only about 1\% of the mechanical wind luminosity needs to be
converted to X-rays, thereby drastically reducing the apparent
mismatch.  At the same time, the flux ratio is predicted one order of
magnitude too high, which means that the temperature needs to come
down from $\sim$500\,MK ($\sim$40\,keV) to $\sim$15\,MK
($\sim$1.3\,keV). Only detailed modelling can show whether the X-ray
properties can be brought into agreement with the observations.

The indications 
of line emission associated with the companion stars of $\gamma$\,Cas
\citep{2002PASP..114.1226M} and $\pi$\,Aqr \citep{2002ApJ...573..812B}
also support the notion of an interaction between the companion stars and the disks of
the Be stars, although it is not clear whether this interaction is
radiative, gravitational or both.  The asymmetric structure in the
disk of BZ\,Cru \citep{2013A&A...550A..65S} may have the same origin.

The BeHeB model reproduces also other observed properties of the $\gamma$\,Cas stars.  
(i) Any interaction between He-star wind and Be disk will lead to a correlation
between X-ray and optical continuum as well as H$\alpha$ line-emission flux.  
(ii) An interaction between He-star wind and Be disk will also often place 
the X-ray-emitting region closer to the B star than accreting-companion models would.  
(iii) Injection of new matter into the disk can easily lead to increased line-of-sight 
column densities of X-ray-attenuating matter.  (iv)  Collision of the He-star wind 
with a Be-star wind strongly structured by co-rotating interaction regions and/or with 
an azimuthally inhomogeneous Be disk often fed by discrete stellar mass-loss 
events may lead to variable X-ray flux on a broad range of time scales.  
Furthermore, the X-ray emission that may arise from two distinct regions in the BeHeB 
model may well relate to the multi-temperature nature of the
observed X-ray continuum in some $\gamma$\,Cas stars (Sect.\,2); 
a similar differentiation was already proposed by \cite{2016ApJ...832..140H}.

\subsection{Further model predictions}

As discussed in Sect.\,\ref{propeller}, the supernova explosions that
ultimately transform BeHeBs with massive helium-star components into
BeXRBs lead to an increase in orbital eccentricity.  By contrast, in
the progenitors of BeXRBs, namely the BeHeBs, the previous
mass-transfer evolution should reduce any earlier eccentricity
to zero, and align the Be spin and the orbital angular-momentum
vector. Therefore, the orbit of the helium star around the Be stars is
expected to be circular and coplanar with the Be disk.  Instead of
being strongly orbitally modulated, as in BeXRBs 
\citep{2001A&A...377..161O}, the X-ray production
in BeHeB should thus be more continuous (but variable due to mass-loss
events from the Be star).  The truncation by the companion 
of the disk \citep{2001A&A...377..161O} would place the locus of formation 
of the X-rays slightly closer to the B star than to the helium star.  

Other predictions resulting from the given ansatz are that
$\gamma$\,Cas stars might have rather massive helium-star
companions. For most of the $\gamma$\,Cas stars, it is not known
whether they are binaries, let alone the masses of any companions.
However, \citet{2012A&A...537A..59N} proposed that the secondary in
$\gamma$\,Cas is a helium star with a mass of about one
1\,M$_{\odot}$, and $\pi\,$Aqr seems to have a companion mass that
does not fit a white dwarf or a neutron star
\citep[2.2-4.5\,M$_{\odot}$][]{2002ApJ...573..812B}.  
\citet{2017A&A...602L...5N} suggested that the companion may be a main
sequence star.  However, this would not explain the Be nature of the
primary, nor the level of the observed X-ray emission. Furthermore,
as $\pi\,$Aqr would be a wide pre-interaction binary in this case,
a circular orbit would be very unlikely.  By contrast, 
the masses of many other sdO stars reported for Be stars are 
far below one solar mass, which may be a challenge as explained above.  


The He-star wind may blow a significant cavity or even a hole into the Be disk. 
While model predictions of this are beyond 
the scope of this study,
it is noteworthy that asymmetries seem to have
been observed in some cases, for instance in BZ\,Cru
(Sect.\,\ref{bzCru}).  Furthermore, the hot helium star can 
locally change the ionisation structure of the disk, leading to 
periodic orbital modulations as already observed in some Be binaries
\citep{2004A&A...427..307R}.  Direct detection of the 
hot helium stars would best be attempted by orbital-phase-resolved 
UV spectroscopy \citep[cf.][]{2013ApJ...765....2P}.

\section{Summary and conclusions}
\label{summary}
Occam's razor offers the insight that the credibility of a proposed
solution to a problem increases with the simplicity of the solution,
where simplicity is often understood as the usage of established
knowledge modules.  While the existence and effects of companion stars
can be addressed observationally (albeit only very tediously for any
given individual Be system), the two magnetic fields of the magnetic
model proposed for $\gamma$\,Cas stars cannot by construction.
Moreover, helium stars are known to exist whereas magnetic fields
caused by subsurface convection or MRI are still awaiting observational
confirmation even in objects they were originally designed for.  On
this ground, Occam would advise to first exhaust the explanatory power
of binary models.

Previous binary models tried to explain the defining X-ray properties
of $\gamma$\,Cas stars in terms of accretion to white dwarfs or
neutron stars.  However, they are struggling in various ways to
reproduce the observations fully.  A generic objection apparently
fuelled by the available observations could be that the release of
gravitational energy that powers such systems can only take place
close to these compact companions, i.e., far away from the B star.
Therefore, the present study has explored a different 
type of companion, namely the short-lived phase of B
stars with a helium-star companion, BeHeBs, which are the progenitors
of BeXRBs, Be$+$WD or Be$+$sdO systems, depending on the mass of the
helium star.

The collision in BeHeBs of a fast stellar wind from a companion with
the Be disk and/or the Be wind as a different concept
has a well-proven analog in the colliding winds of the two
components of massive binary systems.  The discovery of BHeB stars,
i.e.\ of 'normal' B stars with a helium-star companion but without 
circumstellar disk and without $\gamma$\,Cas-like 
X-ray properties, would favour the wind-disk collision 
part of the BeHeB model.  A specific variant of this
idea, namely the interaction of a WD wind with the Be disk, was first
proposed by \citet{2016ApJ...832..140H}.  Because the X-rays do not
have to be generated in the immediate vicinity of the companion,
prospects are much improved that detailed modelling can achieve good
agreement with a wide range of observations.  Closer to the B star and
its mass-loss activity, the door is much wider open towards
reproducing long-term but only little delayed correlations between
variations in X-rays, optical flux, and UV spectral lines.  In
particular, outbursts may well supply the variable amounts of
line-of-sight matter to explain the observed intermittent attenuations
of the soft X-ray flux.

If $\gamma$\,Cas stars do have a companion with such effects, it is
clear that these stars must be of relatively low mass and low optical
luminosity.  With the additional restriction to stars with a strong
wind, only helium stars and WDs remain as candidates.  The helium-star wind model 
has the potential to place the hardness and flux of thermal X-rays 
from $\gamma$\,Cas stars in the
observed domains.  More $\gamma$\,Cas stars should be carefully
screened for hot subluminous companions.  The most conclusive results
can be expected from UV spectroscopy.  

There is no observational evidence of systematic differences 
other than the X-ray properties between
$\gamma$\,Cas stars and the general population of classical Be stars
of the same spectral type.  This agrees with our model 
which is not dependent on any special assumptions about the Be stars themselves other 
than their rapid rotation.  A full reunification of $\gamma$\,Cas and classical
Be stars can be expected from a spectroscopic study of a
representative $\gamma$\,Cas star like that performed by
\citet{2005ApJ...623L.145W} for \object{$\zeta$\,Oph}.  First positive
diagnoses of multimode NRP exist already for $\gamma$\,Cas
(Sect.\,\ref{gCas}) and $\pi$\,Aqr (Sect.\,\ref{piAqr}).  Such work 
may also lead to coarse predictions of mass-loss outbursts of the B
star \citep{1998ASPC..135..343R, 2018pas8.conf...69B} and thereby
facilitate parallel optical and X-ray spectroscopy when the wind of
the helium star interacts with ejecta from the B star. 
A first estimate of the relative X-ray contributions by interactions between 
the He-star wind and the Be disk or the Be wind, respectively, may also 
result.  Combined 
spectroscopy and photometry may provide valuable diagnostics of disk regions not well
probed by other observations which are mostly biased to the denser
parts.  

Because it is difficult to find Be stars that do not pulsate, it may
be feasible (certainly is attractive) to search for any statistical
differences between the pulsation properties of {\it bona fide} single
Be stars and Be stars with different kinds of companion (neutron
stars, WDs, sdO stars, main-sequence stars): Can such a first crude
step towards asteroseismology of Be stars distinguish formation
channels of Be stars?

Furthermore, BeHeB stars should be valuable academies of the short-lived helium stars and their role in the evolution of massive binaries.  An identification of $\gamma$\,Cas stars with these objects would provide the missing link between the unevolved main-sequence binaries and Be binaries with compact companions.

\begin{acknowledgements}
  We thank the referee, Dr. Georges Meynet, for useful comments and suggestions.
  J.B.\ acknowledges support from the FWO Odysseus program under
  project G0F8H6N.  This publication makes use of data
  products from the Wide-field Infrared Survey Explorer (WISE), which
  is a joint project of the University of California, Los Angeles, and
  the Jet Propulsion Laboratory/California Institute of Technology,
  funded by the National Aeronautics and Space Administration.  This
  research has made use of the SIMBAD database
  \citep{2000A&AS..143....9W} and the VizieR catalog access tool
  \citep{2000A&AS..143...23O}, both operated at CDS, Strasbourg,
  France.  This research has made use of NASA's Astrophysics Data
  System (ADS).  
\end{acknowledgements}

\bibliography{dbaade}

\end{document}